\documentclass[a4paper,11pt,oneside, openright]{book}  
\usepackage[toc,page]{appendix}
\usepackage[usenames,dvipsnames]{color} 
\usepackage{titlesec}
\usepackage{graphicx} 
\usepackage{listings} 
\usepackage{courier} 
\usepackage{amsmath}
\usepackage{eso-pic}
\usepackage{amssymb}
\usepackage{amsthm}
\usepackage{color}
\usepackage{epstopdf}
\usepackage{bm}
\usepackage{caption}
\usepackage{subfig}
\usepackage{algorithmic}
\usepackage{algorithm}
\usepackage{xspace}
\usepackage{multirow}
\usepackage{tabularx}
\usepackage{enumerate}

\newcommand{\lqu}{~\lq\lq}
\newcommand{\rqu}{\rq\rq}
\newcommand{\dci}{DCI\xspace}
\newcommand{\tc}{t_c}

\usepackage{epigraph}

\usepackage{hyperref}
\hypersetup{
    colorlinks=true, 
    linktoc=all,     
    linkcolor=blue,  
}

\usepackage[english]{babel}
\usepackage{microtype}
\usepackage[latin1]{inputenc}
\usepackage{fancyhdr}
\usepackage{float}
\usepackage{svg}

\usepackage{wrapfig}
\definecolor{dkgreen}{rgb}{0,0.6,0}
\definecolor{gray}{rgb}{0.5,0.5,0.5}
\definecolor{mauve}{rgb}{0.58,0,0.82}
\numberwithin{equation}{chapter}
\theoremstyle{plain}
\newtheorem{thm}{Theorem}[chapter]

\newtheorem{lem}[thm]{Lemma}
 
\theoremstyle{definition}
\newtheorem{defn}{Definition}[chapter]
\theoremstyle{remark}

\makeatletter
\def\cleardoublepage{\clearpage\if@twoside \ifodd\c@page\else
    \hbox{}
    \vspace*{\fill}
    \vspace{\fill}
    \thispagestyle{empty}
    \newpage
    \if@twocolumn\hbox{}\newpage\fi\fi\fi}
\makeatother

\fancypagestyle{plain}{%
\fancyhead{} 

}

\definecolor{dkgreen}{rgb}{0,0.6,0}
\definecolor{gray}{rgb}{0.5,0.5,0.5}
\definecolor{mauve}{rgb}{0.58,0,0.82}

\addto\captionsenglish{\renewcommand*\abstractname{Abstract}} %

\newcommand\independent{\protect\mathpalette{\protect\independenT}{\perp}}
\def\independenT#1#2{\mathrel{\rlap{$#1#2$}\mkern2mu{#1#2}}}

\titleformat{\chapter}[display]
  {\normalfont\huge\bfseries} 
	{\filleft\Huge\thechapter}{1em}{}

\newcommand{\HRule}{\rule{\linewidth}{0.5mm}}

\begin{document}


\begin{titlepage}
\thispagestyle{empty}
\oddsidemargin=\dimexpr(\oddsidemargin+\evensidemargin)/2\relax

\begin{center}
\vspace*{-3cm}
{\LARGE  Ca' Foscari University of Venice}\\[0.2cm]
\textsc{\Large Department of Environmental \\Sciences, Informatics and Statistics}\\[3cm]
{\LARGE MSc Thesis in Computer Science}\\[1cm]
\HRule \\[0.4cm]
{ \huge \bfseries Functional Dynamical Structures \\ in Complex Systems: \\ an \\  Information-Theoretic Approach  \\[0.4cm] }
\HRule \\[1cm]

{\Large Marco Fiorucci}

\vspace{3cm}

\textsc{\large Supervisor:} \\[0.15cm]
{\Large Prof.~Irene Poli}

\vfill

{\large Academic Year 2014/2015}

\end{center}
\end{titlepage}

\newpage
\thispagestyle{empty}
\null
\newpage

\pagenumbering{roman}
\frontmatter

\thispagestyle{empty}
\null\vspace{\stretch{1}}
\begin{flushright}
{\LARGE \textit{Dedicated to Corinne}}
\end{flushright}
\vspace{\stretch{2}}\null

\newpage

\chapter*{Abstract}
\thispagestyle{empty}
Understanding the dynamical behavior of complex systems is of exceptional relevance in everyday life, from biology to economy. In order to describe the dynamical organization of complex systems, existing methods require the knowledge of the network topology. By contrast, in this thesis we develop a new method based on Information Theory which does not require any topological knowledge. We introduce the Dynamical Cluster Index to detect those groups of system elements which have strong mutual interactions, named as Relevant Subsets. Among them, we identify those which exchange most information with the rest of the system, thus being the most influential for its dynamics. In order to detect such Functional Dynamical Structures, we introduce another information theoretic measure, called D-index. The experimental results make us confident that our method can be effectively used to study both artificial and natural complex systems.




\chapter*{Acknowledgements}
\thispagestyle{empty}
This thesis is the result of the research I performed at the European Centre for Living Technology (ECLT) in Venice under the supervision of Prof. Irene Poli. \\
\indent First of all, I wish to express my sincerest recognition to my supervisor, Prof. Irene Poli, for her continuous guidance and for introducing me in an international highly reputed research institute. \\
\indent During this thesis, I had the opportunity to collaborate with Prof. Roberto Serra and Dr. Marco Villani from the university of Modena and with Dr. Andrea Roli from the university of Bologna. Working with them was particularly fruitful and inspiring.   \\
\indent I enjoyed a lot the pleasant environment in the ECLT with Dr. Matteo Borrotti, Dr. Alessandro Filisetti, Dr. Davide De March, Dr. Debora Slanzi. I would like to thank in particular my tutor, Dr. Alessandro Filisetti, who helped me break through the difficulties in my research and friendly supported me all over the thesis period.  \\
\indent I would like to extend my thanks to Prof. Marcello Pelillo and Prof. Nicoletta Cocco for their valuable discussions and productive collaboration. \\    
\indent More personally, I want to express my gratitude to my family. Since my childhood my parents have always trusted me and stimulated to learn. I'm also grateful to my brothers for all the experiences we have shared together. \\
\indent Finally, a lovely thanks goes to Corinne for her unconditional support and for being a constant source of encouragement during my challenges.

\newpage
\thispagestyle{empty}
\null
\newpage

\begingroup 
  \makeatletter
  \let\ps@plain\ps@empty
  \makeatother
  \tableofcontents
  
	\clearpage
	\listoffigures
	
	\clearpage
	\listoftables   
\endgroup

\mainmatter 

\chapter{Introduction}

\epigraph{The whole is more than the sum of its parts.}%
{Aristotele}

\vspace{0.5cm}

In recent years complex systems attracted increasing interest among scientists of different fields like physics, computer science, chemistry, biology, ecology, economy and social science, just to name a few. This interest arises from the general idea that several natural systems (such as gene regulatory networks, social systems, etc.) and artificial systems (such as economical system, power grid, etc.) are composed of a large number of simple elements which show a collective emergence behavior that cannot be predicted from the description of each particular element. \\
\indent There is no universal definition of complex system, but most researchers in the field would probably agree that it involves numerous components (agents), which may be simple both in terms of their internal characteristics and in the way they interact. These agents exhibit emergence activities, i.e. not derivable from the summations of the activity of the individual components, and a tangled hierarchical self-organization which cannot be understood by referring to simple hierarchical models.\\
\indent Understanding the behavior of complex systems is of exceptional relevance, from both practical and theoretical reasons. This problem has been addressed by researchers which follow two basic approaches. The first one involves the development of mathematical models which abstract the most relevant features of the system to gain scientific insight into its dynamic. The concepts used in the development of such models belong to dynamical system theory, information theory, game theory, network theory, numerical methods and cellular automata. The second approach, instead, relies on the development of a more detailed and realistic models in order to simulate the emergent behavior which appears when the system is observed for a long time and length scales. The tools used in this approach are computational methods, like agent-based simulation and Monte Carlo simulation.\\
\indent In this thesis we study the dynamical organization of complex systems using methods based on information theory. To do so, we model a system using a random field evolving in time. In particular, the dynamical behavior of each system component is described by means of a discrete stochastic process. We observe the system evolution in a given time interval and we estimate the probability distributions of the variables belonging to the random field. Then, exploiting measures based on information theory, which are known as information-theoretic measures, we look for relevant subsets (RSs). A RS is a group of variables which have a strong mutual interaction and that weakly interact with variables which are not in the relevant subset.\\    
\indent In order to find a measure for the interaction among variables, we exploit a fundamental property of the dynamical system from a complex system perspective which was introduced by Kauffman \cite{Kauffman1993}. He stated that there is a continuous exchange of information among the system constituents and also between the environment and the system. \\
\noindent Historically, Tononi and Edelman \cite{Tononi1998} introduced an information-theoretic measure, the Cluster Index, to study biological neural networks which are close to a stationary state. More recently, such Cluster Index was profitably applied to the study of dynamical systems by Villani et al. \cite{Villani2013} and was referred to as Dynamical Cluster Index (DCI).\\
\indent Starting from such ideas, we study the two measures that contribute to the DCI, which are the Integration and the Mutual Information. We show that the analysis of the different parts of the index is extremely useful to better characterize the nature of the identified relevant subsets. In many cases the detected RSs have an intricate nested structure, so that it might not be clear which groups of variables are really important. To overcome this problem, we introduce a sieving procedure to extract only disjoint or partial overlapping RSs \cite{Filisetti2015}.\\		
\indent Furthermore, we extend the DCI introducing causality \cite{Yang2014} in order to investigate the influence of found relevant subsets on the system dynamics. To achieve this goal we use information theoretic concepts like Entropy Rate and Transfer Entropy \cite{Schreiber2000},\cite{Lautier2014}. In particular, on the basis of these measures, we introduce a directionality index \cite{Filisetti2015} to find the direction of information flows which can take place between a RS and other parts of the system. The direction and the amount of information flows are needed to select the leaders, i.e. the most influenced RSs. We call these subsets functional dynamical structures (FDS) to stress their functional role in the system dynamics. \\
\indent Several different application domains are investigated to test the effectiveness and the robustness of this measures. These domains include: random boolean networks, mammalian cell cycle networks and the Mitogen Activated Protein Kinase (MAPK) cascade signaling pathway in eukaryotes.\\   
\indent Finally, it's worth noting that our method does not require any previous knowledge of underlying network topology of the complex system under study, but relies only on the values assumed by the random fields during the observation time interval.\\

\vspace{0.5cm}

\section{Functional Dynamical Structures}


\vspace{0.3cm}

The Dynamical Cluster Index is an extension of a measure introduced by Tononi and Edelman \cite{Tononi1998} to detect clusters in biological neural networks. In particular, Tononi et al. start from the idea that there are functional bounded regions in the brain despite the fact that all brain regions are widespread connected. These brain regions, called functional clusters, are composed of neurons which interact more strongly with neurons belonging to the same region than with neurons belonging to another brain part. The functional aspect of these clusters is suggested by the fact that there is a very fast signal transmission among intra-cluster neurons and they can thus influence many aspects of brain behavior. It's worth noting that this method was applied to networks considering only fluctuations around stable asymptotic state. The time system evolution was indeed not studied. The first generalization of Tononi's method was made by Villani et al. \cite{Villani2013} in order to detect intermediate-level emergent structures in complex systems. They extended the cluster index to the study of truly dynamical systems and the DCI was introduced. \\
\indent In this thesis, we further develop the method based on the Dynamical
Cluster Index and we also show new interesting applications that highlight its effectiveness and help uncovering some of its features \cite{Filisetti2014,Villani2014}. There are of course several methods to identify clusters of nodes in a network based on its topology, but the DCI can be applied also without any prior knowledge on the network topology. This property is shared also by nonlinear correlation methods  \cite{Kai2014}; with respect to the latter, the advantage of the DCI is that it is not limited to binary relations but it can be applied to clusters of any size.
Finally, we study the influence that detected relevant subsets have on the system by focusing on the causal interactions among variables through the different dynamical states. For this purpose, we introduce a directionality index which, as previously mentioned, is an information theoretic measure based on the transfer entropy. To the best of our knowledge, such approach in unprecedented in the study of the dynamical behavior of complex systems.\\

\vspace{0.5cm}

\section{Contributions}

\vspace{0.3cm}

The contribution of this thesis may be summarized as follows:\\

\begin{itemize}
\item we studied the two measures that contribute to the DCI, which are the Integration and the Mutual Information. We showed that the analysis of the different parts of the index is extremely useful to better characterize the nature of the identified relevant subsets. This led to a first publication \cite{Filisetti2014};

\item we extended the Dynamical Cluster Index methodology introducing a sieving algorithm. The results of this extension will appear in \cite{Villani2014}\cite{Filisetti2015};

\item we further extended the DCI methodology introducing causality in order to investigate the influence of found relevant subsets on the system dynamics. The results will be presented at European Conference of Artificial Life \cite{Filisetti2015}.\\
\end{itemize}

\noindent The results obtained in this thesis will be the subject of my oral presentation at the Student Conference on Complexity Science (SCCS2015) Conference in Granada in September 2015.

\vspace{0.5cm}

\section{Outline of the Thesis}       

\vspace{0.3cm}

Chapter 2 reviews the fundamental information theoretic notions like entropy, joint entropy, conditional entropy, relative entropy, mutual information, integration and transfer entropy, which provide a basis for the development of our indexes. Chapter 3 introduces our contribution to the description of complex system behavior, namely the Dynamical Cluster Index, the D-Index and the methodology to detect functional dynamical structures \cite{Filisetti2014,Villani2014,Filisetti2015}. Chapter 4 shows the results of the method application to various models. In particular,  they include what follows: the random boolean network framework (in order to verify that the method is able to identify subsets that make sense) and leaders-followers model (to test the method robustness with respect to increasing noise levels). Furthermore, two applications to real systems are presented, namely mammalian cell cycle networks and the Mitogen Activated Protein Kinase (MAPK) cascade signaling pathway in eukaryotes. In these two latter applications, a peculiar discrete coding which describes the signs of their first-order time derivatives is used. 
In chapter 5, the novelty of the work and the future perspectives are presented.                  
\chapter{Information Theory and Functional Structures}

\epigraph{This word "information" in communications theory relates not so much to what you do say, as to what you could say.}%
{Warren Weaver}

\vspace{0.5cm}

A complex system is composed of a huge amount of interacting parts. The dynamical behavior of the whole system cannot be understood if the analysis just focuses on the properties of the single parts. As a consequence, statistical methods play a key role to understand how the whole system works. In this framework, information theory provides a collection of measures which can be useful to determine the degree of interaction among interconnected elements. In this chapter, we provide a brief background on information-theoretic measures. In particular, we introduce the integration and the transfer entropy which are exploited by our method. In particular, the first one is used to measure the degree of interaction among the system elements while the latter is used to find the direction of the information flow which takes place between two subsets of the system.

\vspace{0.5cm}

\section{Information and Entropy}

\vspace{0.3cm}

Information theoretic measures give us a formal definition of the dynamical structures notion. Even if the concept of information may be referred to different meanings, we consider the formal presentation proposed by Claude Shannon \cite{Shannon1949}, who developed a theory based on a statistic description of a communication system. Shannon generalized the information measure which was presented by Ralph Hartley \cite{Hartley1928}, who stated that the information content of a message of length $n$ composed of symbols chosen from an alphabet of cardinality $N$ is\\  
    
\begin{equation}
I_H(n,N) = nlog_k N
\end{equation}

\vspace{0.35cm}
                                                                                                                                              
\noindent where $k$ is an arbitrary base. Shannon generalized this concept associating a probability distribution to the symbols source, i.e. a symbol from the alphabet is chosen with a certain probability $p$. He stated that the information associated to the observation of a symbol with occurrence probability $p$ is\\  
 
\begin{equation}
I(p) = log \frac{1}{p} 
\end{equation}

\vspace{0.35cm}    
		
\noindent Hence, the information that is received from a symbol observation depends on its occurrence probability, i.e less likely is the symbol appearance and more is the information that is gained from its observation. As a consequence, a source is modeled by a discrete random variable $X$ which can assume value on a non empty finite set $\mathcal{X}$ according to a probability distribution 
$p(x)= Pr\{X=x\}, \, x\in \mathcal{X}$. If the source follows a uniform distribution, the Shannon information coincides with the Hartley information. In order to measure the average amount of transmitted information, Shannon entropy \cite{Shannon1949} was introduced in 1949.

\begin{defn}
The entropy $H(X)$ of a discrete random variable $X$ with alphabet $\mathcal{X}$ and a probability mass function $p(x)= Pr\{X=x\}, \, x\in \mathcal{X}$ is defined as follows \\ 

\begin{equation}
H(X) = - \sum_{x \in \mathcal{X}} p(x) \, logp(x) 
\end{equation}
\end{defn}

\vspace{0.5cm}

\noindent As an alternative notation, we use $H(p)$. We consider the log in base two (bits as units of entropy) and we define $0 \, log0 = 0$(since $x \, logx \rightarrow 0$ as $x \rightarrow 0$). \\
We naturally extend the definition of entropy to a random vector, i.e. a vector composed of two or more random variables.  

\begin{defn}
The entropy $H(\bm{X})$ of a discrete random vector \\ $\bm{X} = (X_1, X_2, \ldots, X_n)$  
with alphabet $\bm{\mathcal{X}} = \mathcal{X}_1 \, \times \, \mathcal{X}_2 \times \, \cdots \times \, \mathcal{X}_n$ and a joint probability distribution $p(\pmb{x})=p(x_1, x_2, \ldots, x_n)$ is defined as follows \\

\begin{equation}
H(\bm{X}) = - \sum_{\pmb{x} \in \bm{\mathcal{X}}} p(\pmb{x}) \, logp(\pmb{x}) 
\end{equation}
\end{defn}

\vspace{1.5cm}

\noindent By analogy with probability theory, we define the conditional entropy using the notion on conditional probability which is a measure of the 
probability of an event given the knowledge of the occurrence of another event. 

\begin{defn}
The conditional entropy of two discrete random variables $X$ and $Y$ with alphabets $\mathcal{X}$ and $\mathcal{Y}$ and a joint 
probability distribution $p(x,y)$ is defined as follows \\

\begin{equation}
H(Y|X) = - \sum_{x \in \mathcal{X}} \sum_{y \in \mathcal{Y}} p(x,y) \, logp(y|x) 
\end{equation}
\end{defn}

\vspace{1.5cm}

\noindent The following theorem describes the relationship between joint entropy and conditional entropy.

\begin{thm}[Chain Rule] 
Given two discrete random variables $X$ and $Y$ with alphabets $\mathcal{X}$ and $\mathcal{Y}$ and a joint probability distribution 
$p(x,y)$, then \\

\begin{equation}\label{eq:chain}
H(X,Y) = H(X) + H(Y|X) = H(Y) + H(X|Y) 
\end{equation}                  
\end{thm}

\vspace{1.5cm}

\noindent We extend the previous result to a generic random vector.

\begin{thm} 
Given a discrete random vector  $\mathbf{X} = (X_1, X_2, \ldots, X_n)$  
with alphabet $\bm{\mathcal{X}} = \mathcal{X}_1 \, \times \, \mathcal{X}_2 \times \, \cdots \times \, \mathcal{X}_n$ and a joint probability distribution $p(\pmb{x})=p(x_1, x_2, \ldots, x_n)$, then \\

\begin{equation}\label{eq:chain_vect}
H(\mathbf{X}) = \sum_{i=1}^n H(X_i|X_{i-1},X_{i-2}, \ldots, X_1)
\end{equation}                  
\end{thm}

\vspace{1.5cm}

\section{A Few Properties of Entropy}

\vspace{0.3cm}

We present few properties of entropy which are exploited to construct our method. In order to prove these properties, we introduce one of the most important inequality in information theory.

\begin{lem}[Gibbs' Inequality] 
Given a random variables $X$ with alphabet $\mathcal{X}$ and two probability mass functions $p(x)\, , q(x) \, , x\in \mathcal{X}$, then \\

\begin{equation}
\sum_{x \in \mathcal{X}} p(x) \, ln\frac{p(x)}{q(x)} \geq 0 
\end{equation}

\vspace{0.3cm}
              
\noindent with equality iff $p(x)=q(x)$, for all $x\in \mathcal{X}$.    
\end{lem}

\vspace{1.5cm}

\noindent Gibbs' inequality allows us to introduce an upper bound for Shannon entropy.

\begin{thm}
Given a discrete random variable $X$ with alphabet $\mathcal{X}$, which contains a number of symbols equals to $|\mathcal{X}|$, and a probability mass function $p(x), \, x\in \mathcal{X}$, then \\ 

\begin{equation}\label{eq:entropy_up}
 0 \leq H(X) \leq log|\mathcal{X}| 
\end{equation}          

\vspace{0.3cm}

\noindent with equality iff $p(x)$ is a uniform distribution. \\

\begin{proof}

$0 \leq p(x) \leq 1$ implies that $log(\frac{1}{p(x)} \geq 0)$. It follows that $H(X) \geq 0$.\\ 
The equality holds iff $\exists x\in \mathcal{X} \, : \, p(x)=1$ because $\forall x\in \mathcal{X}$ \\ $p(x) \, logp(x) = 0 \,  \longleftrightarrow \, p(x) = 0 \, \vee ,\ p(x) = 1  \, \longleftrightarrow ,\ \exists x\in \mathcal{X} \, : \, p(x)=1.$ \\

\vspace{0.3cm}

\noindent From Gibbs' inequality we obtain \\		   

\begin{equation}
H(X) = - \sum_{x \in \mathcal{X}} p(x) \, logp(x) \leq - \sum_{x \in \mathcal{X}} p(x) \, logq(x)  
\end{equation}          

\vspace{0.3cm}

\noindent Let $q(x)$ be uniformly distributed. Hence \\
 
\begin{equation}
- \sum_{x \in \mathcal{X}} p(x) \, logp(x) \leq  \sum_{x \in \mathcal{X}} p(x) \, log|\mathcal{X}| 
= log|\mathcal{X}| \sum_{x \in \mathcal{X}} p(x)  
\end{equation}  

\vspace{0.3cm}

\noindent Since \\

\begin{equation}
\sum_{x \in \mathcal{X}} p(x) = 1
\end{equation}  

\vspace{0.3cm}

\noindent We obtain \\

\begin{equation}
H(X) \leq log|\mathcal{X}|
\end{equation}
 
\vspace{0.3cm}
 
\noindent Finally we prove that if $X$ is uniformly distributed then $H(X)=log|\mathcal{X}|$. \\ 	

\begin{equation}
H(X) = \sum_{x \in \mathcal{X}} \frac{1}{|\mathcal{X}|} \, log|\mathcal{X}| = 
log|\mathcal{X}| \sum_{x \in \mathcal{X}} \frac{1}{|\mathcal{X}|} = log|\mathcal{X}|
\end{equation}  
\end{proof}
\end{thm}

\vspace{1cm}

For our purpose, we interpret the entropy as a measure of a \textit{lack of knowledge} about the system. In particular, we compute the probability of each possible state of a dynamical system after observing its evolution through a certain time interval. The gained information depends on the observed trajectory. Therefore, we can see the entropy as the average information which is gained by the system observation. This approach is borrowed from statistical mechanics, where a macrostate of a system is described associating a probability distribution over the possible microstates which are unknown. In this contest, the entropy is seen as a measure of the disorder of the systems: a connection between the information theory and the thermodynamics can be thus grasped. \cite{Kristian2014}.\\

\vspace{0.5cm}

\section{Mutual Information}

\vspace{0.3cm}

In the study of complex systems, we have to deal with nonlinear systems which do not satisfy the superposition principle. We are thus interested in extensive measures which are useful to characterize a complex systems at least in some particular cases. Intuitively, the information emitted from two independent sources should be equal to the sum of the two individual sources. The log-dependence in the definition of the Shannon entropy makes the entropy of two independent random variables an additive quantity. Hence, the system entropy is maximum when every pair of system elements is independent. Otherwise, the correlation between two or more elements of the systems causes an entropy reduction.\\		 
\indent We introduce mutual information which can be used to measure the degree of dependence of two random variables.

\begin{defn}
The mutual information $I(X,Y)$ of two discrete random variables $X$ and $Y$ with alphabets $\mathcal{X}$ and $\mathcal{Y}$ and a joint probability distribution $p(x,y)$ is defined as follows \\

\begin{equation}\label{eq:MI}
I(X,Y) = \sum_{x \in \mathcal{X}} \sum_{y \in \mathcal{Y}} p(x,y) \, log \frac {p(x,y)}{p(x)p(y)}
\end{equation}
\end{defn}

\vspace{1.5cm} 

\noindent Hence, the mutual information is symmetric under the exchange of its arguments. \\
\indent We present some relationships between entropy, conditional entropy and mutual information. In order to prove these properties, we introduce another extremely important information-theoretic inequality: the Jensen's inequality. We recall some mathematical notions which are used in the inequality definition.

\begin{defn}[Convex function]
Given a function $f: \, [a,b] \rightarrow \mathbb{R}$. $f$ is said to be convex if for all $x_1,x_2 \in [a,b], \, \lambda \in [0,1]$ \\

\begin{equation}
f(\lambda x_1 + (1-\lambda)x_2) \leq \lambda f(x_1) + (1-\lambda)f(x_2)    
\end{equation}
\end{defn}

\vspace{1.5cm} 
   
\begin{thm}
Given $f: \, [a,b] \rightarrow \mathbb{R}$. Let be $f \in \mathcal {C}^1$. $f$ is convex iff   \\

\begin{equation}
f(x) \geq f(x_0) + f'(x_0)(x-x_0) \,\,\, x,x_0\in [a,b]      
\end{equation}
\end{thm}
 
\vspace{1.5cm} 

\begin{defn}[Standard Simplex]
\begin{equation}
\Delta \left \{ \pmb{x} \in \mathbb{R}^n : \sum_{i=1}^{n} x_i=1  \text{ and } \,\,\, x_i \geq 0\, \forall i=1, \ldots, n   \right \}
\end{equation}
\end{defn}

\vspace{1cm}

\begin{figure}[!ht]
	\centering
    \subfloat[]{%
      \includegraphics[scale=0.55]{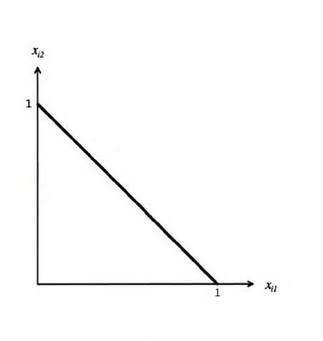}
    }
    \quad\quad\quad
    \subfloat[]{%
      \includegraphics[scale=0.55]{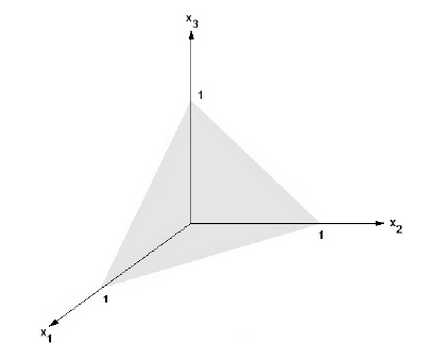}
    }
    \caption{Standard simplex for $n=2$ (a) and $n=3$ (b).}
    \label{fig:simplex}
\end{figure}

\vspace{1cm}

\begin{thm}[Jensen's Inequality]
Given $f: \, [a,b] \rightarrow \mathbb{R}$, $x_1, \ldots, x_k \in [a,b]$, $\pmb{\lambda} = (\lambda_1, \ldots, \lambda_k)\in \Delta_k$. Let be $f$ convex, then \\
  
\begin{equation}\label{eq:jensen}
f \left ( \sum_{i=1}^{k} \lambda_i x_i \right ) \leq \sum_{i=1}^{k} \lambda_i f(x_i)        
\end{equation}

\vspace{0.3cm} 

\noindent whith equality iff $f$ is strictly convex.
\end{thm}

\vspace{1.5cm} 

\noindent Jensen's inequality allows us to introduce a lower bound for mutual information.

\begin{thm}
Let $X$ and $Y$ be two discrete random variables with alphabets $\mathcal{X}$ and $\mathcal{Y}$ and a joint probability distribution $p(x,y)$, then \\

\begin{equation}\label{eq:MI_lower}
I(X,Y) \geq 0 
\end{equation}

\begin{proof}

By definition \\

\begin{equation}
I(X,Y) = \sum_{x \in \mathcal{X}} \sum_{y \in \mathcal{Y}} p(x,y) \, log \frac {p(x,y)}{p(x)p(y)}
\end{equation}

\vspace{0.3cm}

\noindent Since f(z) = log(z) is a concave function, we can apply Jensen's inequality changing the inequality direction \\

\begin{equation}
- I(X,Y) \leq log \sum_{x \in \mathcal{X}} \sum_{y \in \mathcal{Y}} p(x,y) \, \frac {p(x)p(y)}{p(x,y)} 
\end{equation}

\vspace{0.3cm}

\begin{equation}
- I(X,Y)  \leq log \sum_{x \in \mathcal{X}} \sum_{y \in \mathcal{Y}} p(x)p(y)
\end{equation}

\vspace{0.3cm}

\noindent From the normalization axiom of probability theory, it follows \\

\begin{equation}
\sum_{x \in \mathcal{X}} \sum_{y \in \mathcal{Y}} p(x)p(y) = 1
\end{equation}

\vspace{0.3cm}

\noindent Hence \\

\begin{equation}
- I(X,Y)  \leq log 1 = 0  \iff I(X,Y) \geq 0  
\end{equation}
\end{proof}
\end{thm}

\vspace{1cm}

\noindent There are the following relationships between entropy and mutual information which can be proved using \eqref{eq:jensen}, \eqref{eq:chain}. \\       

\begin{equation}
I(X,Y) = H(X) - H(X|Y)   
\end{equation}

\vspace{0.5cm} 

\noindent From \eqref{eq:chain} it follows that \\

\begin{equation}
I(X,Y) = H(X) - H(X|Y) = H(Y) - H(Y|X) = I(Y,X)   
\end{equation}

\vspace{0.5cm} 

\begin{equation}
I(X,Y) = H(X)  + H(Y) - H(X,Y)  
\end{equation}

\vspace{0.5cm} 

\noindent The following theorem states that, on average, the knowledge about a random variable can reduce the uncertainty in another random variable. 

\begin{thm}
Let $X$ and $Y$ be two discrete random variables with alphabets $\mathcal{X}$ and $\mathcal{Y}$ and a joint probability distribution $p(x,y)$, then \\

\begin{equation}\label{eq:conditioning}
H(X|Y) \leq H(X) 
\end{equation}

\vspace{0.3cm}

\noindent with equality iff $X \independent Y$
\end{thm}

\vspace{1cm}

\noindent It's worth nothing that $H(X|Y=y)$ may be greater than $H(X)$. The equation \ref{eq:conditioning} is true only for average quantity, i.e $H(X|Y)$ \cite{Cover}. \\  

\vspace{0.5cm}

The following theorem introduces an upper bound for mutual information. 

\begin{thm}
Let  $X$ and $Y$ be two discrete random variables with alphabets $\mathcal{X}$ and $\mathcal{Y}$ and a joint probability distribution $p(x,y)$, then \\

\begin{equation}\label{eq:MI_upper}
I(X,Y) \leq \min(H(X),H(Y))
\end{equation}

\begin{proof}
Entropy is non negative quantity so, in order to find the maximum of the mutual information, we can find the maximum of the sum H(X)+H(Y) and the minimum of the joint entropy H(X,Y). \\
\noindent From \eqref{eq:chain} we obtain \\

\begin{equation}
H(X,Y)=H(X)+H(Y|X)=H(Y)+H(X|Y)
\end{equation}

\vspace{0.3cm}

\noindent Hence, for every joint distribution p(x,y) holds \\

\begin{equation}
H(X,Y) \geq \max (H(X),H(Y))
\end{equation}

\vspace{0.3cm}

\noindent Hence \\

\begin{equation}
H(X) + H(Y) - H(X,Y) \leq H(X) + H(Y) - \max (H(X),H(Y))
\end{equation}

\vspace{0.3cm}

\noindent Finally we obtain \\

\begin{equation}
I(X,Y) = H(X) + H(Y) - H(X,Y) \leq \min (H(X),H(Y))
\end{equation}

\end{proof}
\end{thm}

Hence, from the equations \ref{eq:MI_lower} and \ref{eq:MI_upper}, we can state that I(X,Y) takes values over the interval $[0,\min(H(X),H(Y))]$.\\ 

\vspace{0.5cm}

\section{Integration}

\vspace{0.3cm}

As we previously hinted at, mutual information can be used to measure the degree of statistical dependence between two elements or subsets of elements when they are associated to two random variables or two random vectors respectively. \\ 
\indent We can generalize this concept in order to measure the degree of interaction among $k$ elements of the system. To do so, let us consider a system $U$ composed of $N$ elements. The system dynamics is described using a random field, i.e. $\pmb{X} = (X_1, X_2, \ldots, X_N)$, where $X_i$ is a discrete random variable associated with the $i$-th element. $X_i$ assumes value in a finite alphabet $\mathcal{X}_i$. Let  $S$ be a set of $k$ elements of the system, such that $k \leq N$, i.e. $S$ is a subset of $U$. Without loss of generality, let $X_1,X_2, \ldots, X_k$ be the random variables associated with the $k$ elements.\\              
If all the random variables are mutually independent, the entropy of the whole subset $S$ can be computed as follows \\

\begin{equation}
H(\mathbf{S}) = \sum_{i=1}^k H(X_i)  
\end{equation}

\vspace{0.5cm}

\noindent If there are interactions among the elements of $S$, then a degree of statistical dependence arises among the $k$ random variables. 
It follows an entropy reduction that can be measured by means of integration which is an information-theoretic measure defined as follows \cite{Tononi1994} \\

\begin{defn}
\begin{equation}
I(\mathbf{S}) = \sum_{i = 1}^k H(X_i) - H(\mathbf{S})    
\end{equation}   
           
\noindent where $\mathbf{S}$ is the random vector composed of all the variables which are in $S$.\\ 
\end{defn}

\vspace{0.3cm}

\noindent The following theorem introduces a lower and an upper bounds for integration.

\begin{thm} 
Given a discrete random vector  $\mathbf{X} = (X_1, X_2, \ldots, X_k)$  
with alphabet $\bm{\mathcal{X}} = \mathcal{X}_1 \, \times \, \mathcal{X}_2 \times \, \cdots \times \, \mathcal{X}_k$ and a joint probability distribution $p(\pmb{x})=p(x_1, x_2, \ldots, x_k)$. Let be $\mathcal{X}_1  = \mathcal{X}_2 = \ldots = \mathcal{X}_k = \mathcal{X}$, then \\ 

\begin{equation}
0 \leq I(\mathbf{X}) \leq (k-1) \, log|\mathcal{X}|
\end{equation}                  
\begin{proof}

\vspace{0.3cm}

\noindent From \ref{eq:chain_vect} we obtain \\

\begin{equation}
H(\mathbf{X}) = \sum_{i=1}^k H(X_i|X_{i-1},X_{i-2}, \ldots, X_1)
\end{equation}  

\vspace{0.3cm}

\noindent From \ref{eq:conditioning} we can state that for $i = 1, \ldots, k$ \\

\begin{equation}
H(X_i|X_{i-1},X_{i-2}, \ldots, X_1) \leq H(X_i,X_{i-1},X_{i-2}, \ldots, X_1)  
\end{equation}  

\vspace{0.3cm}

\noindent Hence \\

\begin{equation}
H(\mathbf{X}) = \sum_{i=1}^k H(X_i|X_{i-1},X_{i-2}, \ldots, X_1) \leq \sum_{i = 1}^k H(X_i) \iff I(\mathbf{X}) \geq 0
\end{equation}  

\vspace{0.3cm}

\noindent To prove that $I(\mathbf{X}) \leq (k-1) \, log|\mathcal{X}|$ we use \ref{eq:entropy_up} \\
 
\begin{equation}
H(X_i) \leq log|\mathcal{X}_i|   \,\,\,\,\,\ i= 1, \ldots, k
\end{equation}  

\vspace{0.3cm} 

\noindent If $X_i$ is uniformly distributed for $i=1,\ldots, k$ then $H(X_i)$ assumes its maximum value (i.e. $log|\mathcal{X}_i|)$. In this case, the minimum value of $H(\mathbf{X})$ is achieved when $\mathbf{X}$ assumes, with probability greater than zero, only $|\mathcal{X}|$ out of the $|\mathcal{X}|^k$ possible values. Under this assumption it follows that $H(\mathbf{X}) \leq  log|\mathcal{X}|$ and hence \\
                       
\begin{equation}
I(\mathbf{S}) = \sum_{i = 1}^k H(X_i) - H(\mathbf{S}) \leq \sum_{i = 1}^k  log|\mathcal{X}| - log|\mathcal{X}| 
\end{equation}  

\vspace{0.3cm} 

\begin{equation}
I(\mathbf{S}) = \sum_{i = 1}^k H(X_i) - H(\mathbf{S}) \leq  klog|\mathcal{X}| - log|\mathcal{X}| = (k-1) \, log|\mathcal{X}|  
\end{equation}  
\end{proof}
\end{thm}

\vspace{1cm}

\begin{figure}[!ht]
	\centering
    \includegraphics[scale=0.55]{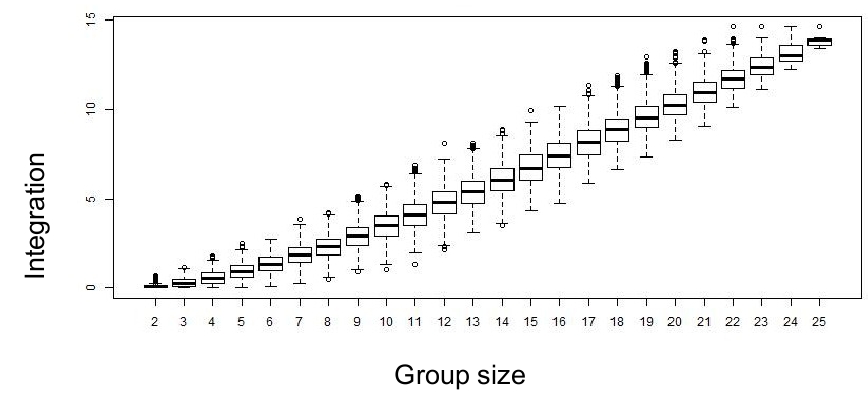}
    \caption{Integration scaling with respect to the subset size. The analyzed system is a catalytic reaction networks \cite{Villani2014} composed of $26$ elements.}
    \label{fig:I_scaling}
\end{figure}

\begin{figure}[!ht]
	\centering
	\vspace{1.5cm}
    \includegraphics[scale=0.55]{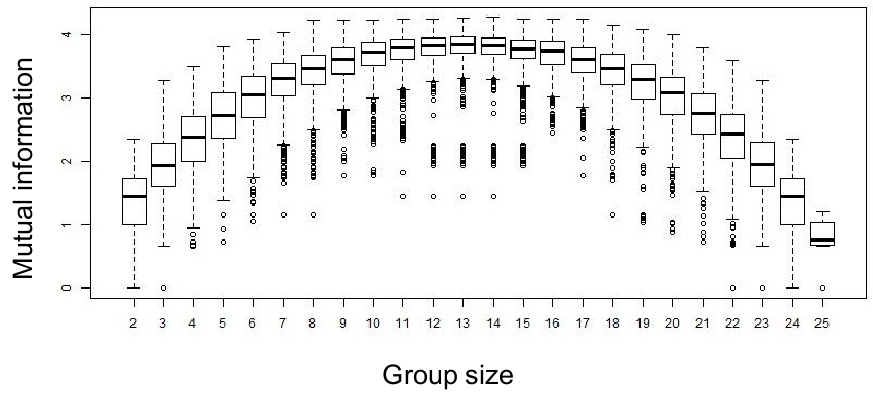}
    \caption{Mutual Information scaling with respect to the subset size. The analyzed system is a catalytic reaction networks \cite{Villani2014} composed of $26$ elements.}
    \label{fig:MI_scaling}
		\vspace{1cm}
\end{figure}

For all of these features we can use integration and mutual information to detect relevant subsets (RS) in complex dynamical systems. As we previously hinted at, a RS is a subset of state variables which heavily influence the system dynamics. It is composed of variables which have a strong mutual interaction and that weakly interact with variables which do not belong to the subset. Integration can be used as internal criterion: it is a measure of the internal interaction among the elements of a RS. At the same time, mutual information can be used as an external criterion: it is a measure of the interaction between the RS and the rest of the system (or between two relevant subsets). A candidate relevant subset is characterized by a high integration value and a low mutual information value. \\
It's worth noting that both integration and mutual information scale with the size of a relevant subset. We can see this behavior in \ref{fig:I_scaling}, \ref{fig:MI_scaling}, where integration and mutual information values are computed analyzing the dynamics of a catalytic reaction network composed of 26 molecules (see chapter $5$ for more details). In order to compare RSs of different sizes, a normalization procedure has been developed. It will be presented in the next chapter.\\


\vspace{0.5cm}
 
\section{Entropy Rate and Transfer Entropy}

\vspace{0.3cm}

As previously explained, mutual information can be used to measure the interaction between two relevant subsets. Since mutual information is a symmetric measure, it does not provide indications on the information flows directions. Furthermore, it does not take into account dynamical information which are needed to detect the subsets which most influence the system dynamics. We say that a subset $A$ affects the behavior of a subset $B$ if the net information flow goes from $A$ to $B$. Hence, the direction and the amount of information flows is needed to select the leaders, i.e. the most influential RSs. We call these subsets Functional Dynamical Structures (FDS) to stress their functional role in the system dynamics. A FDS sends a high amount of information to the other subsets conditioning their behavior. In order to measure the dynamical information flow between two relevant subsets, we used a new measure based on the transfer entropy \cite{Schreiber2000}. We recall some notions from stochastic  processes theory \cite{Bass2011} and we define the entropy rate which is used in the transfer entropy definition.\\
        
\begin{defn}[Stochastic Process]
A stochastic process $\{X_i\}$ is a sequence of discrete random variables with alphabet $\bm{\mathcal{X}} = \mathcal{X}^k$ and a joint probability distribution $p(\pmb{x})=p(x_1, x_2, \ldots, x_n)$ for $n=1,2,\ldots$.  
\end{defn}

\vspace{0.3cm}

\begin{defn}[Markov Process]
A stochastic process $\{X_i\}$ is called a Markov Process or Markov Chain if     
\begin{equation}
Pr\{X_{n+1}= x_{n+1} | X_n= x_n, \ldots, X_1= x_1\} = Pr\{X_{n+1}= x_{n+1} | X_n= x_n\}
\end{equation}
for $n=1,2,\ldots$ and for all $x_1, \ldots, x_{n+1}$.
\end{defn}

\vspace{0.3cm}

\begin{defn}[Stationary Markov Process]
A Markov Process $\{X_i\}$ is stationary if \\     

\begin{equation}
Pr\{ X_{n+1}= k | X_n= l\} = Pr \{X_2= k | X_2 = l\}
\end{equation}

\vspace{0.3cm}

\noindent i.e. $p(x_{n+1}|x_n)$ is independent with respect to $n$.\\
\end{defn}

\vspace{1cm}

In order to introduce the notion of entropy rate, let us consider a system $U$ composed of $N$ elements. The system dynamics is described using a random field evolving in time, i.e. $\pmb{X}(t) = (X_{1,t}, X_{2,t}, \ldots, X_{N,t})$, where $X_{i,t}$ is a discrete random variable associated with the $i$-th element at time $t$. $X_i$ assumes value in a finite alphabet $\mathcal{X}_i$. Hence, if we fix $t$, $\pmb{X}$ is a random field while, if we fix $i$, $X_i(t)$ is a stochastic process. Let assume that $X_i(t)$ is a stationary Markov process of order k.
Under this assumption, the probability to find $X_i$ in state $x_i$ at time $t+1$ satisfies the following relation \\

\begin{equation}
p(x_{i,t+1}|x_{i,t},\ldots,x_{i,t-k+1}) = p(x_{i,t+1}|x_{i,t},\ldots,x_{i,t-k}) 
\end{equation}

\vspace{0.5cm}

\noindent We adopt the following shorthand notation proposed by Schreiber \cite{Schreiber2000} \\

\begin{equation}
x_{i,t}^k = (x_{i,t},\ldots, x_{i,t-k+1}) 
\end{equation}

\vspace{0.3cm}

\noindent for words of length $k$.

\vspace{1.5cm}

\noindent Let us now define the entropy rate as follows

\begin{defn}[entropy rate]
Given a stationary Markov process $\{X_i\}$ of order $k$, the entropy rate $h_{X_i}$ is \\      

\begin{equation}
h_{X_i} = \sum_{\mathcal{X}_i} p(x_{i,t+1},x_{i,t}^k) \, log \, p(x_{i,t+1}|x_{i,t}^k) 
\end{equation}

\vspace{0.3cm}

\noindent for $n=1,2,\ldots$ and for all $x_1, \ldots, x_{n+1}$.
\end{defn}
 
\vspace{1.5cm}

\noindent Since $p(x_{i,t+1}|x_{i,t}^k) = p(x_{i,t+1}^{k+1}/ p(x_{i,t}^k)$, we can express $h_{X_i}$ as follows\\

\begin{equation}
h_{X_i} = H_{X_i^{k+1}} - H_{X_i^k}
\end{equation}
   
\vspace{0.3cm}

\noindent where  $H_{X_i^{k+1}}$ and $H_{X_i^{k}}$ are the entropies of the process ${X_i}$ considering two delay vectors of dimension $k+1$ and $k$ respectively.

\vspace{1.5cm}

\noindent In order to find the direction of the information flow which takes place between two elements $X_i$ and $X_j$, we have to generalize the entropy rate. The idea is to measure the deviation from the following generalized Markov property \\

\begin{equation}
p(x_{i,t+1}|x_{i,t}^k) = p(x_{i,t+1}|x_{i,t}^k, x_{j,t}^l)
\end{equation}

\vspace{1.5cm}

\noindent If there is an information flow from $X_j$ to $X_i$, then the transition probabilities of $X_i$ depends on the states of $X_{j,t}$. In this case the entropy rate is \\    

\begin{equation}
h_{1,X_i} = \sum_{\mathcal{X}_i}\sum_{\mathcal{X}_j} p(x_{i,t+1}, x_{i,t}^k, x_{j,t}^l)	\, log \, p(x_{i,t+1}|x_{i,t}^k,x_{j,t}^l) 
\end{equation}

\vspace{1.5cm}

\noindent Otherwise, if there is not an information flow from $X_j$ to $X_i$, then the transition probabilities of $X_i$ are independent from the state of $X_j$. In this case the entropy rate becomes \\
 
\begin{equation}
h_{2,X_i} = \sum_{\mathcal{X}_i}\sum_{\mathcal{X}_j} p(x_{i,t+1}, x_{i,t}^k, x_{j,t}^l) \, log \, p(x_{i,t+1}|x_{i,t}^k) 
\end{equation}

\vspace{1.5cm}

\noindent In order to measure the degree of statistical dependence of the transitional probabilities of $X_i$ on that of $X_j$, we can use the transfer entropy which is defined as follows \\

\begin{equation}
\begin{split}
T_{X_j\rightarrow X_i}& = h_{2,X_i} - h_{1,X_i}=\\
& =\sum_{X_i,X_j} p(x_{i,t+1}, x_{i,t}^k, x_{j,t}^l) \, log \, \frac{p(x_{i,t+1}|x_{i,t}^k,x_{j,t}^l)}{ p(x_{i,t+1}|x_{i,t}^k)}
\end{split}
\end{equation}

\vspace{1.5cm}

\noindent At the same time, we can define the transfer entropy from $X_i$ to $X_j$ as follows \\

\begin{equation}
T_{X_i\rightarrow X_j} = \sum_{X_i,X_j} p(x_{j,t+1}, x_{j,t}^k, x_{i,t}^l) \, log \, \frac{p(x_{j,t+1}|x_{j,t}^k,x_{i,t}^l)}{ p(x_{j,t+1}|x_{j,t}^k)}
\end{equation}

\vspace{1.5cm}
 
\noindent We can express transfer entropy in terms of conditional entropies as follows \\

\begin{equation}
T_{X_j\rightarrow X_i} = H(X_{i,t+1}|X_{i,t}^k) - H(X_{i,t+1}|X_{i,t}^k, X_{j,t}^l)  
\end{equation}

\vspace{1cm}

\begin{equation}
T_{X_i\rightarrow X_j} = H(X_{j,t+1}|X_{j,t}^k) - H(X_{j,t+1}|X_{j,t}^k, X_{i,t}^l)  
\end{equation}

\vspace{1.5cm}

\noindent It's worth noting that the transfer entropy is explicitly non-symmetric: $T_{X_j\rightarrow X_i}$ is in general different from $T_{X_i\rightarrow X_j}$. In the next chapter, we present an index to detect the directionality of information flow between two relevant subsets.

\chapter{Dynamical Cluster Index \\ Methodology}

\epigraph{An idea which can be used once is a trick. If it can be used
more than once it becomes a method.}%
{G. Polya and S. Szeg\"{o}}

\vspace{0.5cm}

\section{The Dynamical Cluster Index}

\vspace{0.5cm}

The \dci is an extension of the Functional Cluster Index (CI) introduced by Edelman and Tononi in 1998 \cite{Tononi1998} and aimed at detecting functional clusters in brain regions. In our work \cite{Villani2014}, we relax the stationary constraint and extend the CI to actual dynamical systems, so as to apply it to a wide range of system classes, from abstract models to biological models.\\
As in our work we rely on observational data, probabilities are estimated by the relative frequencies of the values observed for each variable. \\
Let us now consider a system $U$ composed of $K$ variables (e.g. agents, chemicals, genes, artificial entities) and suppose that $S_{k}$ is a subset composed of $k$ elements, with $k<K$. The value $DCI(S_{k})$ is defined as the ratio between two aforementioned measures: the integration ($I$) and the Mutual Information ($MI$). \\ As previously stated, $I(S_{k})$ measures the statistical independence of the 
$k$ elements in $S_{k}$ and is defined as:\\

\begin{equation}
	\label{eq:i}
	I(S_{k})~=\sum_{s \in S_{k}} H(s)-H(S_{k})
\end{equation}

\vspace{1cm}

\noindent As previously stated, mutual information $MI(S_{k};U\backslash S_{k})$ measures the mutual dependence between subset $S_{k}$ and the rest of the system $U\backslash S_{k}$ and it is defined as: \\

\begin{equation}
	\label{eq:mi}
	MI(S_{k};U\backslash S_{k})~= H(S_{k}) + H(U\backslash S_{k}) - H(S_{k},U\backslash S_{k})
\end{equation}

\vspace{0.5cm}

\noindent hence, the  \dci,\\

 \begin{equation}
	\label{eq:dci}
	DCI(S_{k})~=\frac{I(S_{k})}{MI(S_{k};U\backslash S_{k})}
\end{equation}

\vspace{1cm}

\noindent The value of \dci is not defined if $MI(S_{k};U\backslash S_{k})=0$. However, a vanishing mutual information is a sign of separation of the investigated subset from the rest of the system, and therefore the subset must be studied separately.
It is worth noting that the \dci scales with the subset size. In \cite{Villani2014} we show a procedure to normalize it, nevertheless a better approach is that of assessing the statistical significance of the \dci of $S_{k}$ by means of a statistical significance index \cite{Tononi1998}:\\

 \begin{equation}
	\label{eq:tc}
	\tc(S_{k}) ~= \frac{DCI(S_{k}) - \left\langle DCI_{h} \right\rangle}{\sigma(DCI_{h})}
	~=\frac{\nu DCI-\nu\left\langle DCI_{h} \right\rangle}{\nu\sigma(DCI_{h})}
\end{equation}

\vspace{0.5cm}

\noindent where $\left\langle DCI_{h} \right\rangle$ and $\sigma(DCI_{h})$ are respectively the average and the standard deviation of the $DCI$ of a sample of subsets of size $k$ extracted from a reference system $U_{h}$ randomly generated according to the frequency of each single state in $U$ and $\nu=\left\langle MI_{h} \right\rangle / \left\langle I_{h} \right\rangle$ is the normalization constant. It is worth noting that the aim of the reference system is that of quantify the finite size effects affecting the information theoretical measures on a random instance of a system with finite dimensions.\\

\vspace{0.5cm}

\section{DCI: remarks and observations}

\vspace{0.3cm}

As previously stated, the two factors used to compute the $DCI$ of a candidate RS are the integration of its elements and the mutual information between the subset and the rest of the system. 
Therefore, the $DCI$ of a subset $S$ of variables in a system---i.e., a possible candidate RS---is then estimated by collecting a set of system states and then by computing the entropy values of the possible combinations of variables in $S$. Therefore, all we need is just a collection of observations of the system and, in principle, no information on the topology or on the internal mechanics of the system are required.
The rationale behind the ratio between $I$ and $MI$ is that a candidate RS should express higher interactions among its components than with the rest of the system. In spite of the simplicity of the concept, one must carefully consider the meaning of the two quantities and the way they are combined. In fact, the implicit assumption here is that it holds $MI>0$ and that the orders of magnitude of $I$ and $MI$ are reasonably close. However, there might be cases in which $MI\approx0\; \wedge\; I>0$, which denotes integrated subsets dynamically independent from the rest of the system. Furthermore, we may be interested in finding the most integrated subsystems among the ones that exchange information with the rest of the system. Moreover, it is important to remark that insightful information can be also provided by analyzing the two factors composing the $DCI$, before computing their ratio. For this reason, an assessment of the information brought by the individual values of $I$ and $MI$ can shed light on the potential of the application of these information-theoretic measures to detect candidate RSs.\\

\vspace{0.5cm}

\section{Integration vs. mutual information}
\label{sec:I-vs-MI}

\vspace{0.3cm}

\begin{figure}[tp]
\begin{center}
 \includegraphics[scale=0.4]{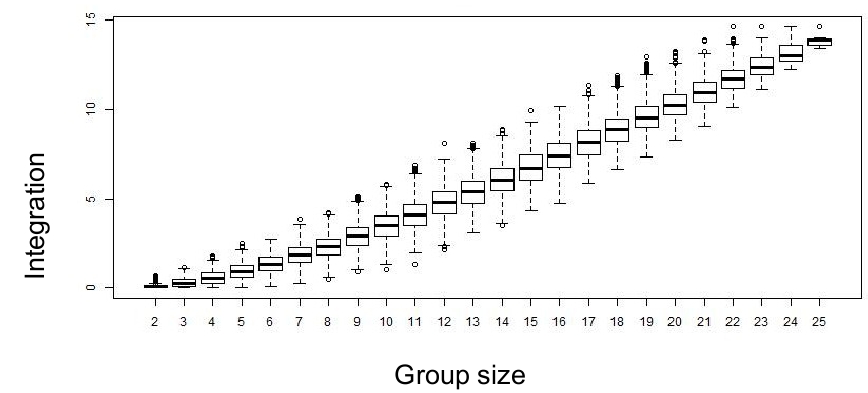}
\end{center}
\caption{Typical statistical behaviour of integration for different subsystem sizes. The system analyzed is a catalytic reaction system \cite{Villani2013} composed of 26 molecular species.}
\label{fig:integration}
\end{figure}

\begin{figure}[tp]
\begin{center}
 \includegraphics[scale=0.4]{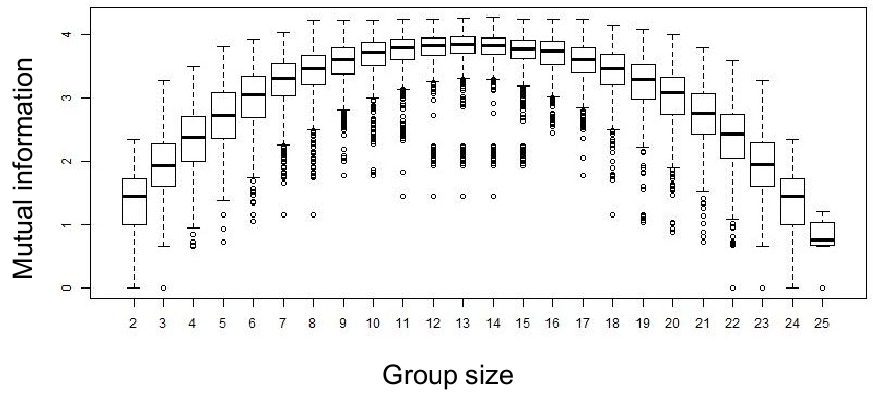}
\end{center}
\caption{Typical statistical behavior of mutual information for different subsystem sizes. The system analysed is a catalytic reaction system~\cite{md-ecal2013} composed of 26 molecular species.}
\label{fig:mutual-info}
\end{figure}

\noindent To assess the contribution of $I$ and $MI$ we first study their statistical behavior with respect to subsystem size. We analyzed several dynamical systems and the results obtained show that the general trend respects the theoretical findings \cite{Tononi1998}, as shown in Figure~\ref{fig:integration} and Figure~\ref{fig:mutual-info}. 

As we can observe, $I$ scales approximately linearly, whilst $MI$ is non-monotonic. We have previously proved that the maximal value of $I$ for a subset of size $n$ equals $n-1$. Hence, we can define a rescaled version of $I$, $rI = I/n-1$, useful to compare the integration of subsets of different dimensions.
Moreover, the addition of a variable $x$ to the considered subset $S$ increases $I$ by 1 if $x$ deterministically depends  on any variable in $S$, while it leaves $I$ unchanged if it assumes random values. These properties may be useful to reckon the relative importance of integration values computed for subsystems of different size.\\

\vspace{0.5cm}

\subsection{Experiments on a tuneable model}

\vspace{0.3cm}

\begin{figure}[t]
\begin{center}
 \includegraphics[scale=0.4]{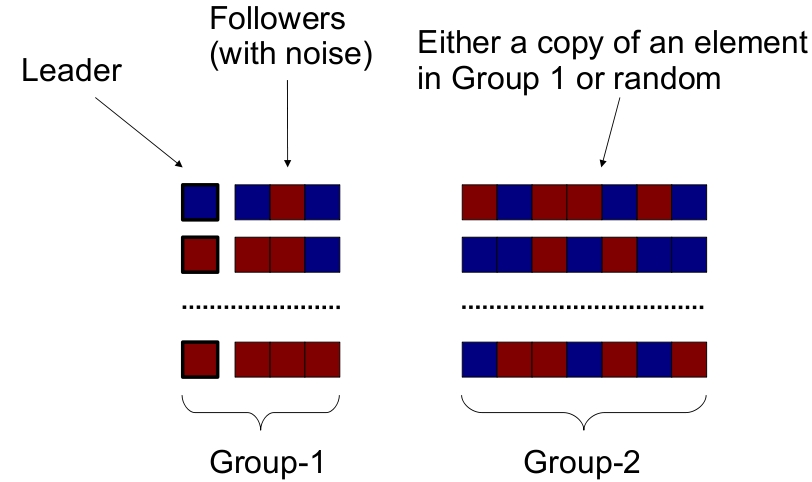}
\end{center}
\caption{Data generated according to a simple leader-followers model.}
\label{fig:leader-followers}
\end{figure}

We are interested in understanding the individual informative contribution of $I$ and $MI$. To this aim, we study these measures on a simple model in which the integration among variables in a subsystem under observation and its mutual information can be tuned by acting on two parameters. The model abstracts from specific functional relations among elements of the system and could resemble a basic leader-followers model (see Figure~\ref{fig:leader-followers}). The system is composed of a vector of $n$ binary variables $x_1,x_2,\ldots,x_n$, e.g., representing the opinion in favor or against a given proposal. The model generates independent observations of the system state, i.e. each observation is a binary $n$-vector generated independently of the others, on the basis of the following rules:

\begin{itemize}
 \item Variables are divided into two groups, $G_1 = [x_1,\ldots,x_k]$ and $G_2 = [x_{k+1},\ldots,x_n]$;
  \item $x_1$ is called the {\em leader} and it is assigned a random value in $\{0,1\}$;
  \item the value of the {\em followers} $x_2,\ldots,x_k$ is set as a copy of $x_1$ with probability $1-p_{noise}$ and randomly with probability $p_{noise}$;
  \item the values of elements of $G_2$ are assigned as a copy of a random element in $G_1$ with probability $p_{copy}$, or a random value with probability $1-p_{copy}$.
\end{itemize}

\noindent It is possible to tune the integration among elements in $G_1$ and the mutual information between $G_1$ and $G_2$ by changing $p_{noise}$ and $p_{copy}$. Note that, given significant level of integration, we have two notable cases:
\begin{itemize}
 \item[{\em(a)}] $MI \approx 0 \rightarrow$ isolated (possibly integrated) RS;
  \item[{\em(b)}] $MI \gg 0 \rightarrow$ integrated and segregated cluster.
\end{itemize}

\begin{figure}[tp]
\begin{center}
 \includegraphics[scale=0.4]{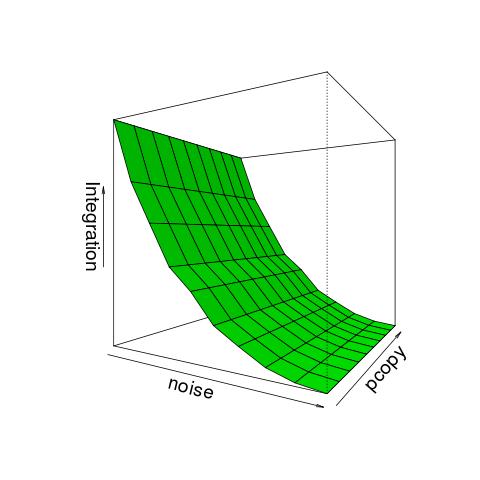}
\end{center}
\caption{Integration of $G_1$ as a function of $p_{noise}$ and $p_{copy}$.}
\label{fig:plot3d-integration}
\end{figure}
\begin{figure}[tp]
\begin{center}
 \includegraphics[scale=0.4]{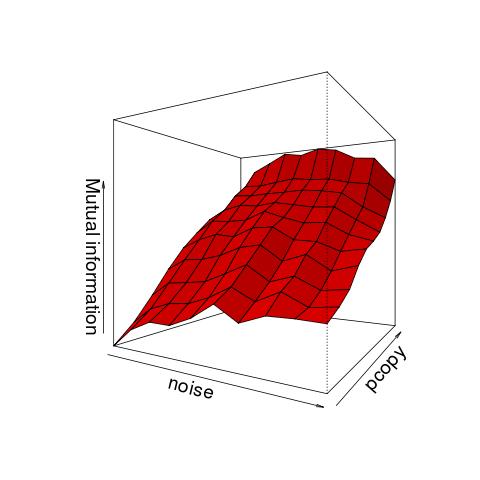}
\end{center}
\caption{Mutual information between $G_1$ and $G_2$ as a function of $p_{noise}$ and $p_{copy}$.}
\label{fig:plot3d-mutual_information}
\end{figure}

\noindent The possible scenarios which can be obtained by tuning $p_{noise}$ and $p_{copy}$ can be conveniently illustrated by a 3-dimensional plot. Figure~\ref{fig:plot3d-integration} shows the behavior of integration of $G_1$ as a function of $p_{noise}$ and $p_{copy}$. We can observe that it is a decreasing function of $p_{noise}$, while it is independent of $p_{copy}$ (by definition, indeed). The behavior of the mutual information between $G_1$ and $G_2$ is depicted in Figure~\ref{fig:plot3d-mutual_information}. As we can observe, $MI$ increases fast with $p_{copy}$, as this parameters increases the correlation between variables in $G_2$ and $G_1$. Moreover, it also increases with $p_{noise}$, but the reason is that the correlation among variables in $G_1$ increases the randomness of variables in $G_1$, which behaves similarly to the variables in $G_2$.

The case of $MI \approx 0$ corresponds to the situation in which $G_1$ is almost completely independent of $G_2$ and can be easily detected by observing only $MI$.
Conversely, if we are interested in discovering $G_1$ as significant RS, then we would consider cases in which both $I$ and $MI$ are significantly high. In our experiments, these cases correspond to $p_{noise} < 0.3$ and $0 < p_{copy} < 0.5$. In these scenarios, we find that it is possible to detect $G_1$ by using only $I$, divided by $n-1$. It is important to mention that, when $p_{noise}$ is slightly higher, it is necessary to resort to the computation of the DCI to detect group $G_1$.  

\vspace{0.5cm}

\section{The Relevant Subset Detection Algorithm}

\vspace{0.3cm}

As we previously hinted at, the dynamical cluster index is used to detect the Candidate Relevant Subsets (CRSs) of a system. In order to develop a methodology to detect CRSs, let us consider a system $U$ composed of $N$ elements. The system dynamics is described using a random field, i.e. $\pmb{X} = (X_1, X_2, \ldots, X_N)$, where $X_i$ is a discrete random variable associated with the $i$-th element. $X_i$ assumes value in a finite alphabet $\mathcal{X}_i$. Let  $S$ be a set of $k$ elements of the system, such that $k \leq N$, i.e. $S$ is a subset of $U$.
In order to get a list of Candidate Relevant Subsets (CRSs), we compute the $DCI$ of every possible subset of variables in $U$ and ranking the subsets by $DCI$ values. The subsets occupying the first positions are most likely to play a relevant role in system dynamics. For large-size systems, exhaustive enumeration is computationally impractical as it requires to enumerate the power set of $U$. In this case, we resort to a genetic algorithms. \\
The list of CRSs can be ranked according to the significance of their $DCI$. We can directly use this ranking to identify by hand the relevant CRSs for the dynamics of the system. Nevertheless, in many cases this analysis might return a huge list of entangled CRS, so that a direct inspection is required for explaining their relevance.
To this aim, we present a \dci analysis post-processing sieving algorithm to reduce the overall number of CRS to manually tackle.  
The main assumption of the algorithm is that if a CRS $A$ is a proper subset of CRS $B$, i.e. $A \subset B$, then only the subset with the higher DCI is maintained between the two. Thus, only disjoint or partially overlapping CRSs are retained: the used assumption implies that the remaining CRSs are not further decomposable, forming in such a way the\lqu building blocks\rqu~of the dynamical organization of the system. The pseudo-code of the algorithm is presented in Algorithm~\ref{alg:sieve}. \\

\vspace{0.5cm}

\begin{algorithm}
\caption{Sieving algorithm}
\label{alg:sieve}
\begin{algorithmic}
  \STATE \textbf{Input:} The array $C$ of all the $CRS$ ranked by their $\tc$(\dci)
  \STATE \textbf{Output:} A subset $RS \subseteq CRS$
  \STATE $RS \leftarrow \emptyset$
  \STATE $n \leftarrow |CRS|$
  \STATE Initialize auxiliary array $Del[k] \leftarrow \{0 \; {\rm for} \; k \; {\rm in} \; 1 \ldots n\}$
  \FOR{$i = 1$ to $n-1$}
    \FOR{$j = i+1$ to $n$}
      \IF {$Del[i] \neq 1 \;\; \wedge \;\; Del[j] \neq 1$}
	\IF {$C[i] \subset C[j] \;\; \vee \;\; C[j] \subset C[i]$}
	  \STATE $Del[j] \leftarrow 1$
	\ENDIF
      \ENDIF
    \ENDFOR
  \ENDFOR
  \FOR{$i = 1$ to $n$}
    \IF {$D[i] = 0$}
      \STATE $RS \leftarrow RS \cup \{C[i]\}$
    \ENDIF
  \ENDFOR
\end{algorithmic}
\end{algorithm}

\vspace{0.5cm}

\section{Temporal Correlation: the D Index}

\vspace{0.3cm}

Although by the application of the \dci, CRSs are detected, this measure does not provide indications neither on the quantity nor on the direction of the information which flow among subsets. 
To this aim we apply the directionality index proposed in \cite{Lautier2014}.

Let $X$ and $Y$ be two random variables---or, equivalently, two sets of variables. As mentioned in the previous chapter, we can define the entropy rate of $X$ as the average number of bits needed to encode a successive state of $X$ if all the previous states are known, considering that the value of $X$ at $t+1$ depends either on $X$ and $Y$ at the time $t$, eq.~\ref{eq:h1}, or just on the value of $X$ at the time $t$, eq.~\ref{eq:h2}.\footnote{Note that the temporal dependency is not necessarily of unitary lag, i.e. $t-1 \rightarrow t$. For a complete assessment of the statistical dependency of $X$ on $Y$ one should sum over $t-1, t-2, \ldots, t-T$, where $T$ is the observation time limit. Nevertheless, note that (i) in this work we are analyzing Markovian systems, whose behavior depends only from the immediately previous step and (ii) although TE is not a direct measure of causal effect, the use of short history length alters the character of the measure towards inferring causal effect~\cite{Lizier2010}.} \\

\begin{subequations}
\begin{align}
	\label{eq:h1}
	h_{1} &=-\sum_{X,Y} p(x^{t+1},x^{t},y^{t})log~p(x^{t+1}|x^{t},y^{t})\\
	\label{eq:h2}
	h_{2} &=-\sum_{X,Y} p(x^{t+1},x^{t}+,y^{t})log~p(x^{t+1}|x^{t})
\end{align}
\end{subequations}

\noindent then, the transfer entropy $T$ is defined as the difference between the aforementioned entropy rates. $T$ describes how the uncertainty of $X$ is reduced by knowing the previous states of $Y$ and $X$ itself, eq.~\ref{eq:te}.\\

\begin{equation}
\label{eq:te}
\begin{split}
T_{Y\rightarrow X}& =h_{2}-h_{1}\\
& =\sum_{X,Y} p(x^{t+1},x^{t},y^{t})~log~\frac{p(x^{t+1}|x^{t},y^{t})}{p(x^{t+1}|x^{t})},
\end{split}
\end{equation}

\noindent Thus, $T_{Y\rightarrow X}$ can be described in term of entropy as:

\begin{equation}
\label{eq:texy}
T_{Y\rightarrow X}=H(X^{t+1}|X^{t})-H(X^{t+1}|Y^{t},X^{t}),
\end{equation}

\noindent and, since the $T_{Y\rightarrow X}$ describes the information moving from $Y$ to $X$, and the transfer entropy is not symmetric, the information from $X$ to $Y$ is computed as well, eq.~\ref{eq:teyx}. 

\begin{eqnarray}
\label{eq:teyx}
T_{X\rightarrow Y}=H(Y^{t+1}|Y)-H(Y^{t+1}|X^{t},Y^{t}),
\end{eqnarray}

\noindent Once that $T_{Y\rightarrow X}$ and $T_{X\rightarrow Y}$ are known, the directionality $D$ of the information flow between $X$ and $Y$ can be measured as:

\begin{eqnarray}
\label{eq:d}
    D_{X\rightarrow Y}~= 
	\begin{cases}
    		0 ,& \text{if } T_{X\rightarrow Y} = T_{Y\rightarrow X}\\
    		\frac{T_{X\rightarrow Y} - T_{Y\rightarrow X}}{T_{X\rightarrow Y} + T_{Y\rightarrow X}} \in [-1,1],   & \text{otherwise}
\end{cases}
\end{eqnarray}

\noindent where $D_{X\rightarrow Y}=1$ indicates that all the information moves from $X$ to $Y$, i.e. absence of information flow from $Y$ to $X$ and, conversely, $D_{X\rightarrow Y}=-1$ indicates that $X$, with respect to $Y$, is just an information receiver and not an information sender.\\
It is worthwhile to notice that $D$ does not provide any indication about the amount of information exchanged between the variables, but it only provides suitable indications on the direction of the information flow. \\

\chapter{Experimental Results}

\epigraph{It doesn't matter how beautiful your theory is, it doesn't matter how smart you are. 
If it doesn't agree with experiment, it's wrong.}%
{Richard P. Feynman}

\vspace{0.5cm}
 
\section{Boolean Networks}

\vspace{0.3cm}

We consider a system composed of $12$ boolean nodes updated on the basis of either a boolean function or a random boolean value generator. Nodes update their state in parallel and synchronously. We illustrate the results of $5$ instances of this network, defined in Table~\ref{tab:boolnetw} \footnote{Note that the size of these systems allows for an exhaustive enumeration of all the possible groups, allowing their complete assessment. It is worth remarking that each perturbation is introduced after the system has recovered a stable dynamical situation.}.
\noindent The $5$ instances share a common structure but differ in specific dynamical organizations of some nodes. 

\begin{figure*}
\begin{center}
\includegraphics[width=2in]{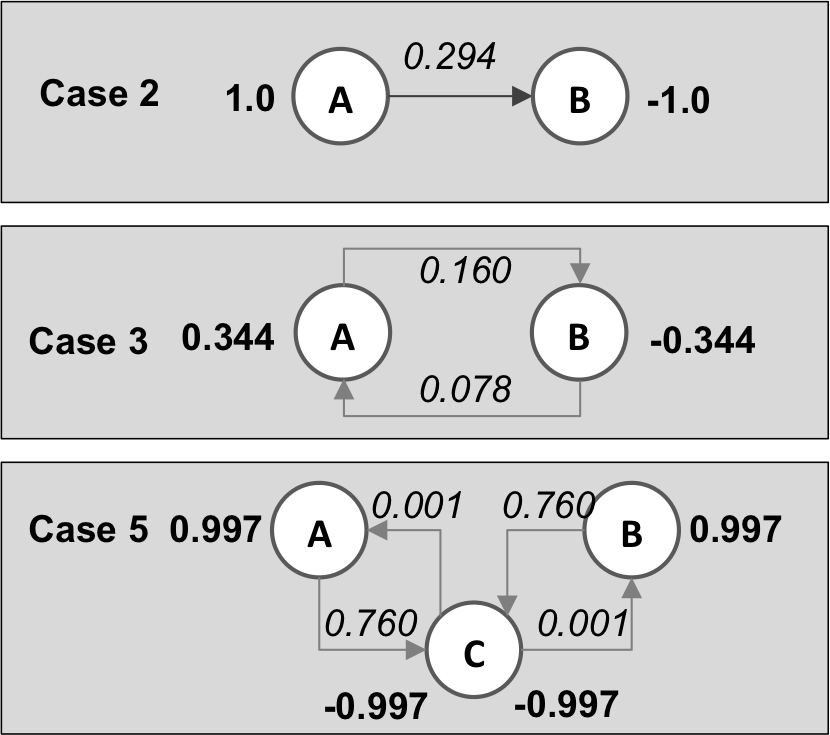}(a)
\includegraphics[width=2.3in]{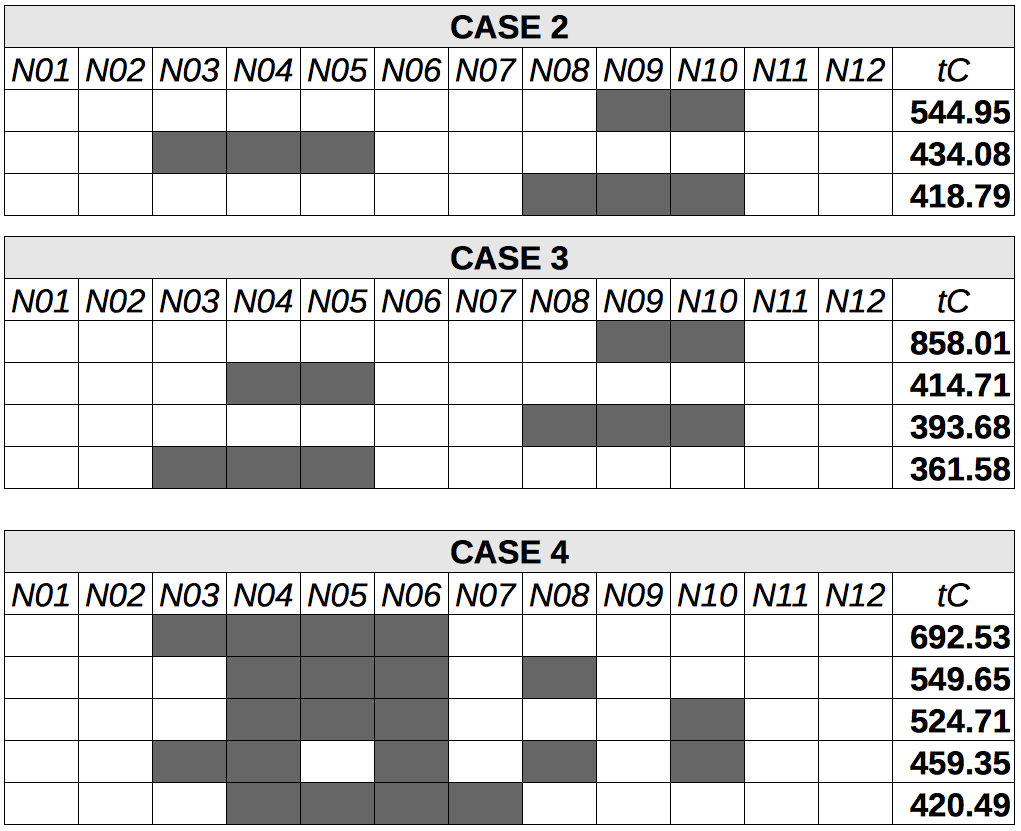}(b)
\caption{(a) The dynamical organization of the systems described in the text; on the edges, the Transfer Entropy, and on the nodes the $D$ index. \textit{Group C} in \textit{case 5} has two values for the $D$ index, each value depicting the \textit{group C} role in the relation with \textit{group A} and \textit{group B}, respectively. (b) The CRS identified by the \dci only in some systems described in the text. Each row denotes one CRS, composed of nodes whose entries are marked in black; on the right, the value of $\tc$.
Note that in \textit{case 2} the third row is the correct identification of \textit{group B}, and that in \textit{case 3} the third and fourth rows are \textit{group A} and \textit{group B}.  For \textit{case 4} the output of the combination of the \dci analysis and sieving algorithm are presented; note that besides the correct group formed by nodes \textit{N03--N06} other CRS have high $\tc$ values, highlighting the presence of a larger but less evident dynamical group. }
\label{fig:cases}
\end{center}
\end{figure*}

\begin{itemize}

\item In \textit{case 1}, we consider two integrated groups of three nodes (namely, group A and group B), by assigning at each node the $XOR$ function of the other two nodes in the group. In this case the seaving algorithm filtered the $94\%$ of the evaluated CRS, making it possible to easily identify the subsets responsible for the dynamical organization of the system.
Only the two correct CRSs have high $\tc$ values, whereas all the other ones have $\tc$ values lower of more than 2 orders of magnitude. Moreover, no information exchange takes place between group A and group B, as they are structurally independent.\\

\item \textit{Case 2} derives from case $1$ by introducing in the first node of group B a further dependence from the last node of group A, hence introducing information transfer from group A to group B. The combination of the \dci analysis and the sieving algorithm correctly recalls the dynamical organization of the system---i.e. group A influences the behavior of (a part of) group B. 
We observe that the whole group B (Figure~\ref{fig:cases}b) was anyway ranked high w.r.t. its \dci significance, but it was discarded by the sieving algorithm because the dynamics of its first node is influenced also by group A and so the assessment of the whole group B is weakened. In general, the amount of this difference depends on the strength of the forces that influence the interface nodes and the elements interfacing different CRS can be detected by a simple comparison between the \dci analysis and the sieving algorithm outputs.\\

\item \textit{Case 3} derives from \textit{case 2} by introducing a further dependence of the first node of group A from the last node of group B: again, the combination of the \dci and the sieving algorithm detects the interface nodes and the underlining dynamical system organization. Note that the asymmetry in transfer entropies (and therefore in $D$ index) are due to differences in the initial conditions of the boolean network trajectories: a shift in initial conditions can change the direction of this asymmetry.\\

\item In \textit{case 4} five heterogeneously linked nodes are influenced by a triplet that is identical to that of group A. The combination of the dynamical rules of the nodes and their initial condition makes the dynamical behavior of the sixth node always in phase with the triplet, so our analysis individuates this quartet as the most relevant CRS. The other dynamical relations are not sufficiently strong to coordinate all the $8$ nodes, nevertheless their influence creates some masks\footnote{CRSs can be represented by rows, where entries corresponding to variables belonging to CRS are marked in black (\lqu masks\rqu in the following)} having high $\tc$ values, partially overlapped with the leading quartet. The overlap of these masks indirectly indicates the presence of a greater group with respect to the winning quartet, but having a less evident dynamical presence (see Figure~\ref{fig:cases}b).\\

\item \textit{Case 5} derives from \textit{case 1} by adding two nodes whose dynamical behavior directly depends on nodes of both group A and group B: these 8 nodes form therefore a group clearly different from the remaining 4 random nodes, as they are interdependent and ruled by deterministic functions.  This group is identified by the plain  \dci method, but the combination of \dci and sieving algorithm strikingly enlightens the interpretation of two triplets directly influencing a couple of nodes (Figure \ref{fig:cases}a).\\

\end{itemize}

\begin{table*}
\center
\scalebox{0.65}{
\begin{tabular}{|c|c|c|c|c|c|}\hline
Node & \multicolumn{5}{|c|}{Node Rule} \\ \hline\hline
& \textbf{Case 1} & \textbf{Case 2} & \textbf{Case 3} & \textbf{Case 4} & \textbf{Case 5}\\
N01 & Random($0.5$) & Random($0.5$) & Random($0.5$) & Random($0.5$) & Random($0.5$)\\
N02 & Random($0.5$) & Random($0.5$) & Random($0.5$) & Random($0.5$) & Random($0.5$)\\
N03 & (N04~$\oplus$~N05) & (N04~$\oplus$~N05) & N10$\wedge$(N04~$\oplus$~N05) & (N04~$\oplus$~N05) & (N04~$\oplus$~N05)\\
N04 & (N03~$\oplus$~N05) & (N03~$\oplus$~N05) & (N03~$\oplus$~N05) & (N03~$\oplus$~N05) & (N03~$\oplus$~N05)\\
N05 & (N03~$\oplus$~N04) & (N03~$\oplus$~N04) & (N03~$\oplus$~N04) & (N03~$\oplus$~N04) & (N03~$\oplus$~N04)\\
N06 & Random($0.5$) & Random($0.5$) & Random($0.5$) & (N05~$\oplus$~N08) & (N05~$\oplus$~N08)\\
N07 & Random($0.5$) & Random($0.5$) & Random($0.5$) & (N07$+$N08$+$N09$+$N10) $\geq 2$ & $\neg$(N05~$\oplus$~N08)\\
N08 & (N09~$\oplus$~N10) & N05$\wedge$(N09~$\oplus$~N10) & N05$\wedge$(N09~$\oplus$~N10) & N03~$\oplus$~N05 & N09~$\oplus$~N10\\
N09 & (N08~$\oplus$~N10) & (N08~$\oplus$~N10) & (N08~$\oplus$~N10) & (N04$+$N05$+$N07$+$N08)$\leq 2$& (N08~$\oplus$~N10)\\
N10 & (N08~$\oplus$~N09) & (N08~$\oplus$~N09) & (N08~$\oplus$~N09) & N06$\wedge$(N05~$\oplus$~N09) & (N08~$\oplus$~N09)\\
N11 & Random($0.5$) & Random($0.5$) & Random($0.5$) & Random($0.5$) & Random($0.5$)\\
N12 & Random($0.5$) & Random($0.5$) & Random($0.5$) & Random($0.5$) & Random($0.5$) \\ \hline
\end{tabular}
}
\vskip 0.25cm
\caption{The update rules of the boolean networks discussed on the text. Random($0.5$) denotes a Bernoulli distribution with probability $0.5$.} 
\label{tab:boolnetw}
\end{table*}

\vspace{0.5cm}

\section{Gene regulatory networks}

\vspace{0.3cm}

In this section we show the application of our method to two models of regulatory networks: a model of mammalian cell cycle network (MCC in the following), as~\lq\lq booleanized\rq\rq~in~\cite{Faure2006}---see Table~\ref{tab:gene} for the chosen boolean model---, and a model of one of the major cellular signal transduction pathways, known as the Mitogen Activated Protein Kinase (MAPK) cascade~\cite{Widmann1999}. \\

\vspace{0.3cm}

 \subsection{Mammalian cell cycle networks}

\vspace{0.3cm}

\begin{figure*}
\begin{center}
\includegraphics[scale=0.2]{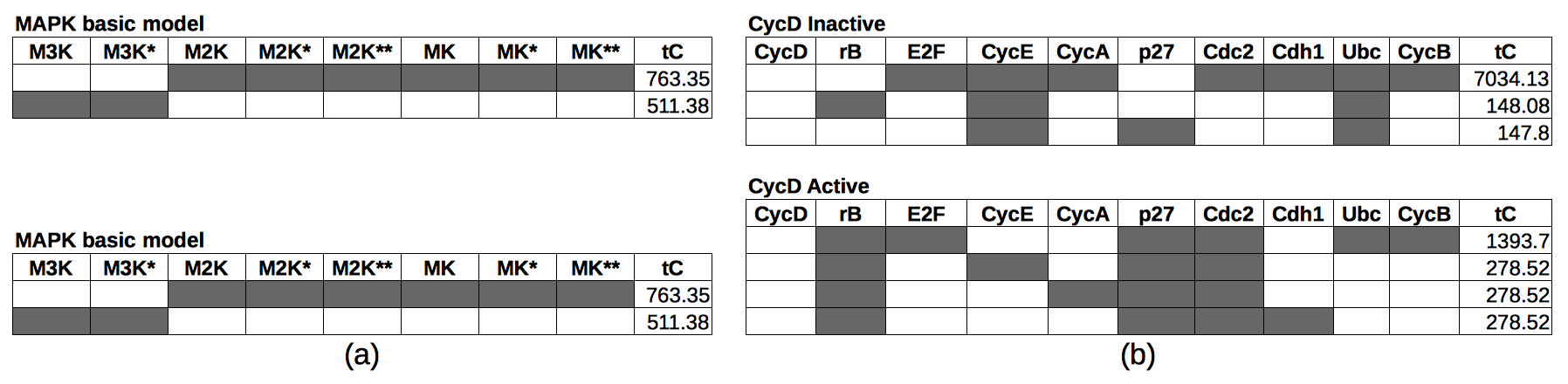}
\caption{(a) The masks identify by the combination of the \dci and the sieving algorithm for (a) the Mitogen Activated Protein Kinase (MAPK) cascade and (b) the two dynamical conditions of the mammalian cell cycle network. In each situation only the masks having significantly high $tC$ values are represented.}
\label{fig:bio}
\end{center}
\end{figure*}

In~\cite{Faure2006} the authors provides a boolean dynamical model of the mammalian cell cycle, able to reproduce the main characteristics of the succession of molecular events leading to the reproduction of the genome of a cell and its division into two daughter cells.\\
Mammalian cell division must be coordinated with the overall growth of the organism; this coordination is achieved through extra-cellular signals whose balance decides whether a cell will divide or remain in a resting state. The positive signals or growth factors ultimately elicit the activation of Cyclin D (CycD) in the cell. In the proposed model CycD thus represents the input and its activity is considered constant. By pointing the interested reader to~\cite{Faure2006} for the details, for now it is enough to say that in absence of CycD the system presents a unique stable attractor where only Rb, p27 and Cdh1 are active, whereas in its presence E2F, CycE, CycA, Cdc20, Cdh1, UbcH10 and CycB cycle with a period of length 7. We perturb both asymptotic states, obtaining in each case only one group (composed of E2F, CycE, CycA, Cdc20, Cdh1, UbcH10 and CycB in the first case, and of Rb, E2F, p27, Cdc20, UbcH10 and CycB in the second case). The leading groups of the two situations are different, but in each case the other CRS identified overlap with these groups and their sum cover the whole system, indicating the presence of a single coordinated pattern. So, the analysis indicates that the elements of the mammalian cell cycle network act as a single compact engine, see~Figure\ref{fig:bio}b.

\begin{table*}
\center
\scalebox{0.6}{
\begin{tabular}{|l|l|l|}\hline
Product & Logical Rule leading to an activity of product & Legend \\ \hline\hline
$CycD$ & $CycD$ & Cyclin D\\
$Rb$ & $(\overline{CycD}\wedge\overline{CycE}\wedge\overline{CycA}\wedge\overline{CycB})\vee(p27\wedge\overline{CycD}\wedge\overline{CycB})$ & Retinoblastoma Protein\\
$E2F$ & $(\overline{rB}\wedge\overline{CycA}\wedge\overline{CycB})\vee(p27\wedge\overline{rB}\wedge\overline{CycB})$ & Transcription factors\\
$CycE$ & $(E2F\wedge\overline{rB})$ & Cyclin E\\
$CycA$ & $(E2F\wedge\overline{rB}\wedge\overline{Cdc20}\wedge\overline{Cdh1}\wedge\overline{Ubc})\vee(CycA\wedge\overline{RB}\wedge\overline{Cdc20}\wedge\overline{Cdh1}\wedge\overline{Ubc})$ & Cyclin A\\
$p27$ & $(\overline{CycD}\wedge\overline{CycE}\wedge\overline{CycA}\wedge\overline{CycB})\vee(p27\wedge\overline{CycE}\wedge\overline{CycA}\wedge\overline{CycB}\wedge\overline{CycD})$ & $p27$ enzyme inhibitor\\
$Cdc20$ & $CycB$ & \multirow{2}{32ex}{Activators of the  Anaphase Promoting Complex} \\
$Cdh1$ & $(\overline{CycA}\wedge\overline{CycB})\vee(Cdc20)\vee(p27\wedge\overline{CycB})$ & \\
$Ubc$ & $(\overline{Cdh1})\vee(Cdh1\wedge Ubc\wedge(Cdc20\vee CycA\vee CycB))$ & E2 ubiquitin conjungating enzyme \\
$CycB$ & $(\overline{Cdc20}\wedge\overline{Cdh1})$& Cyclin B\\ \hline
\end{tabular}
}
\caption{Adapted from~\cite{Faure2006}, the boolean regulatory network of mammalian cell cycle network and a short description of each node of the system.}
\label{tab:gene}
\end{table*}

\vspace{0.5cm}

\subsection{Matabolic pathway MAPK}
\label{sec:mapk}

\vspace{0.3cm}

\begin{figure*}
\begin{center}
\includegraphics[scale=0.3]{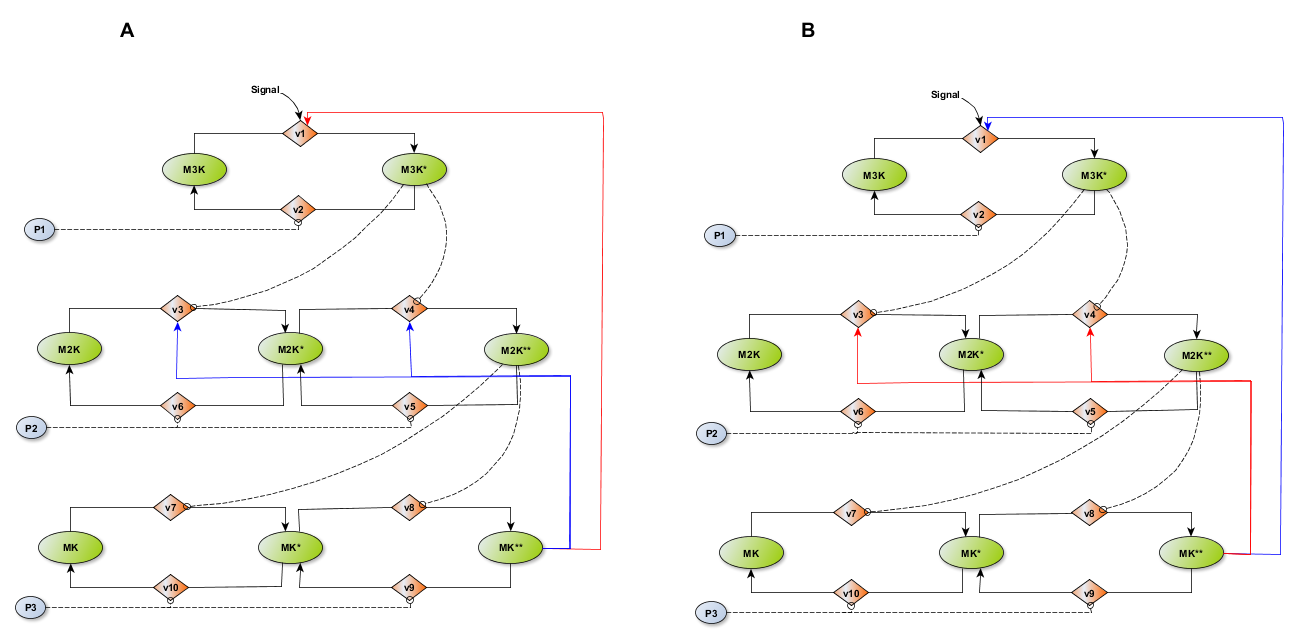}
\caption{(A) Basic model: the scheme of the three layers MAPK cascade reaction pathway is represented (\lqu*\rqu~stands for the phosphorylation). The signal catalyzes the phosphorylation of M3K to M3K* that in turn catalyzes the phosphorylation of M2K to M2K* and the successive phosphorylation of M2K* to M2K**. Finally M2K** performs the double phosphorylation of MK in MK** that is the final output of the MAPK cascade. P1, P2 and P3 dephosphorylate M3K, M2K and MK kinases respectively. V1-V10 stand for the involved reactions. Dashed lines with circle head represent catalysis; the figure highlights the presence of the three~\lqu layers\rqu~described on the text. (B) Two distinct positive and negative feedbacks are added to the basic model: the negative feedback goes from MK** to the second layer (M2K, M2K* and M2K**) while the positive feedback goes from MK** to the first layer (M3, M3K*).}
\label{fig:mapk}
\end{center}
\end{figure*}

The MAPK pathway (evolutionarily conserved from yeasts to humans) responds to a wide range of external stimuli, triggering growth, cell division and proliferation~\cite{Sarma2012}. \cite{Sarma2012} also introduce the models considered in our analysis. The basic model is composed of reactions that are quite well-established from an experimental viewpoint, and it has the hierarchical structure shown in Figure~\ref{fig:mapk}a. The three chemicals identified as the core of this three-layered system are the $MAPKKK$, $MAPKK$ and $MAPK$ kinases (respectively $M3K$ , $M2K$  and $MK$ for short)~\cite{Widmann1999}. $M3K$ is activated by means of a single phosphorylation whereas $M2K$ and $MK$ are both activated by double phosphorylation. Parallel to the phosphorylation by kinases, phosphatases present in the cellular volume can dephosphorylate the phosphorylated kinases (Figure~\ref{fig:mapk}a shows the schema of the $MAPK$ cascade where each layer of the cascade is dephosphorylated by a specific phosphatase). Note that superimposed on the three-layered structure of substrates-product reactions there is the properly called $MAPK$ signalling cascade, a linear chain of catalysis (dashed lines in Figure~\ref{fig:mapk}a)  that transmit the external signal from $M3K^{*}$ to $MK^{**}$.\footnote{The symbol\lqu$*$\rqu~indicates the phophorylated version of the molecule.}  

\begin{figure}
\begin{center}
\includegraphics[scale=0.5]{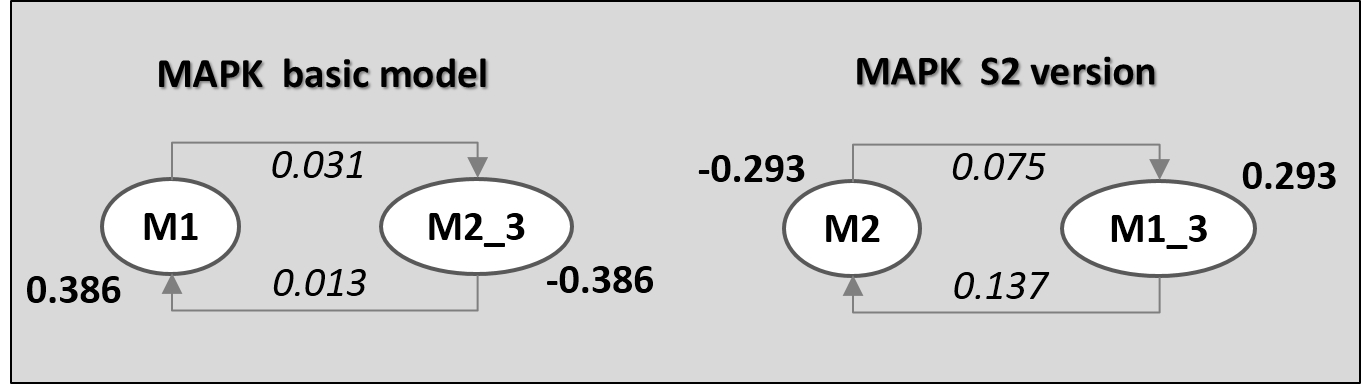}
\caption{The dynamical organization of MAPK system (basic model and S2 version). In italic the Transfer Entropy, in bold the $D$ index associated to the interested group within the relation. M1 group involves the first layer of~\ref{fig:mapk}a; M2 group involves the second layer, whereas groups M2\_3 and M1\_3 involve respectively the layers 2 and 3, and 1 and 3.}
\label{fig:mapkmasks}
\end{center}
\end{figure}

\noindent When the external signal and the concentrations of the phosphatases are kept constant, a top-down reaction scheme as the one described in Figure~\ref{fig:mapk}a leads to fixed-point solutions. On the other hand, oscillations have been reported in the MAPK cascade~\cite{Shin2009} and, in order to account for them,~\cite{Sarma2012} adopt a models with feedback, one of which is described in Figure~\ref{fig:mapk}b. This variant (called S2 in the following) is characterized by a configuration of the activating and inhibiting interactions among layers that alters the~\lqu layered\rqu~structure of the basic model, which is no longer strictly hierarchical.
This alternative model is grounded on experimental data; we will not enter here a discussion about the merits and limits of the model, referring the interested reader to the original paper, but we will take it\lqu for granted\rqu~and we will apply our method to test whether it can discover significant dynamical organization, without any prior knowledge of the interactions, but on the sole basis of the dynamics of concentrations.  We simulate the~\cite{Sarma2012} models with the CellDesigner software~\cite{Funahashi2008,Funahashi2003}, keeping the P1, P2 and P3 phosphatases as constant (as they do) obtaining the same asymptotic states shown by the authors. 
In order to apply our method we perturb the asymptotic states of these models: in particular, we focus our analysis on kinases perturbations. In particular, we perform $10$ perturbation cycles, each cycle involving the perturbation of each single kinase and the successive relaxing to a stable situation before the subsequent perturbation \cite{Villani2014}. The stable situations that are reached can show both oscillating (S2 system) or constant concentrations (basic system). Concentration changes are more significant than their absolute values (it is important to monitor the variables that are changing in coordinate way); therefore the continuous concentration values are represented according to a three levels code related to the sign of the time derivatives at time $t$ (\lqu decreasing concentration\rqu,~\lqu no significant change\rqu,~\lqu increasing concentration\rqu). The combination of \dci and sieving algorithm applied to the basic MAPK model detects two dynamical groups: the first including the first layer of Figure~\ref{fig:mapk}a and the second including the other two layers). The two groups exchange information, the second transmitting more information to the first one (see Figure~\ref{fig:mapkmasks}). The introduction of the feedbacks changes system dynamics: there are still two dynamically relevant groups, now composed of the second layer and by the other two layers, respectively. The analysis therefore suggests that the MAPK system may be decomposed in two dynamically distinct parts.

\chapter{Conclusion}

\epigraph{This is not the end. It is not even the beginning of the end. But it is, perhaps, the end of the beginning.}%
{Winston Churchill}

In this thesis we introduced a methodology based on information theory to identify functional dynamical structures in complex systems. To do so we modeled a system using a random field evolving in time. In particular, the dynamical behavior of each system component is described by means of a discrete stochastic process. We observed the system evolution in a given time interval and we estimated the probability distributions of the variables belonging to the random field. Then, exploiting information-theoretic measures (Dynamical Cluster Index and D-index), we looked for functional dynamical structures.  \\
The method does not require any previous knowledge of underlying network topology of the complex system, but relies only on the values assumed by the random fields during the observation time interval.\\
The effectiveness of the methodology has been validated on test cases and subsequently applied to two prominent biological models, i.e. the mammalian cell cycle network and Mitogen Activated Protein Kinase (MAPK) cascade.\\

\begin{enumerate}[(a)]
\item Random boolean networks.
\item	Mammalian cell cycle networks.
\item	Mitogen Activated Protein Kinase (MAPK) cascade.
\end{enumerate}

\vspace{0.5cm}

\section{Contributions and Novelty of the Work}

\vspace{0.3cm}

The main novelty of the thesis, in comparison to previous application of the cluster index and of similar measures \cite{Tononi1998} is that we use it to consider truly dynamical systems, and not only fluctuations around stable asymptotic states. In principle, different kinds of data can be considered. In the case of a deterministic dynamical system, attractors are the main candidates to provide the required time series and we have shown in case (a) that they can work effectively. Note however that the method is ineffective in a situation that is sometimes encountered, i.e. if there is just a single fixed point. Therefore we conclude that the use of attractor states is only effective if the attractor landscape is rich enough to show the main features of the system organization. This has usually to be evaluated a posteriori, with the exception of trivial cases like that of a single fixed point.\\
It is worth noting that the three-level coding used in case (c) regards the similarities of the derivatives rather than those of the values of the variables and that the models that were used in this work to generate the time series are all based on first-order ordinary differential or difference equations; it should be verified whether the approach is valid also when higher-order dynamical systems are considered. Of course, high-order ODE systems can be transformed in first-order systems by adding variables, but the auxiliary variables that are required might turn out to be unobservable.\\
When a system is subject to continuous external disturbances the time series directly provide the required data and the experimental results show that our treatment can reveal its organizational features even when a high noise level is present.\\
Actually, the range of applicability of this method is quite broad and it does not necessarily need to be limited to dynamical system. Indeed the method just needs a set of frequencies of co-occurences of the values of the system variables. So the method can be applied also to many other systems, since all that is required is a series of "cases" associated to vectors of numerical variables that are not necessarily ordered in time (think for example of different patients, each one described by a vector of values of various symptoms).\\
A final comment is that the method is not a brute-force one: when there is a clear organization in the system, e.g. in case (a), the organization can be read out directly from the order list of candidate relevant sets, but this does not always happen. The study of case (b) and (c) shows that even in entangled real systems the method provides useful clues to uncover the system organization. In order to describe the dynamics of these entangled systems, we introduce a sieving algorithm which selects disjoint or partially overlapping CRSs with the most high DCI values and we used the D-index in order to recover the direction of information flow among these CRSs.\\

\vspace{0.5cm}

\section{Future Work}

\vspace{0.3cm}

As far as future directions of this thesis are concerned, let us only mention some major ones. Several aspects have been mentioned above and will be subject to further analysis, including the effective analysis of high-order dynamical systems with a simplified discrete coding and the comparison among different kinds of time series data, i.e. attractors, transients from arbitrary initial conditions and perturbations.\\
Our method needs discrete variables, so when dealing with continuous data we have introduced a three-level coding that necessarily misses some information. Different coding schemes could be compared, moreover one might also consider the continuous generalization of the proposed method.\\
Other research will be devoted to improvements of the method: it is apparent that it faces a huge computational problem for large systems, since the number of possible subsets increases extremely fast with the number of variables. We used a genetic algorithm to efficiently search the optimal candidate relevant subsets in terms of Dynamical Cluster Index. However, the evaluation of the fitness function can be very expensive when high dimensional systems are considered. In order to deal with high dimensionality, a more efficient fitness function needs to be developed and other heuristic algorithms need to be tested.\\
Several other specific cases need to be studied to confirm the validity of the approach, mainly from real-world data. Let us mention that, among others, we are considering applications to the study of innovation processes, where data come from the real world and not from models. In this respect, it is important to realize that the DCI methodology may be integrated with other approaches: when dealing with real-world problems and not from data generated by a perfectly known model, a typical situation that is often encountered is that one knows some relationships between variables, but not them all. In this case, the most promising approach would be based on a combination of the a priori knowledge with the DCI, in ways that still need to be tested.\\
Moreover, it should be recalled that the Dynamical Cluster Index is just one out of several information-theoretic measures that might be applied to analyze dynamical systems. It is worth noting that the integration and the mutual Information can be useful even if used in isolation and not combined together in the DCI. But other measures, e.g. those that refer to joint distributions at different times, might prove to be particularly useful for the study of dynamical systems.\\
Finally, it is worth noting that the information exchange assessment is just preliminary; we are now working on a method to normalize and then to evaluate the significance of the information transmitted among CRS.\\

\nocite{*}\
\bibliography{References}
\bibliographystyle{abbrv}
\end{document}